\documentclass[
  aps,
  prl,
  preprint,
  floatfix,
  superscriptaddress,
  showkeys,
  amsmath,
  amssymb,
  longbibliography
]{revtex4-2}

\usepackage[english]{babel}
\usepackage{amsmath,amssymb,bbm,mathrsfs,bm,braket,color,graphicx,comment,amsfonts,dsfont}
\usepackage[mathscr]{euscript}
\usepackage{xcolor}
\usepackage{ulem}
\usepackage{orcidlink}
\usepackage{multirow}
\usepackage{color,soul}
\usepackage{siunitx}
\usepackage{lineno}

\begin{document}

\title{Orientation Dependent Resistivity Scaling in Mesoscopic NbP Crystals}

\author{Gianluca Mariani} \thanks{These two authors contributed equally} \affiliation{IBM Research Europe - Zurich, 8803 Ruschlikon, Switzerland}
\author{Federico Balduini} \thanks{These two authors contributed equally}  \affiliation{IBM Research Europe - Zurich, 8803 Ruschlikon, Switzerland}
\author{Nathan Drucker}\affiliation{IBM Research Europe - Zurich, 8803 Ruschlikon, Switzerland}
\author{Lorenzo Rocchino}\affiliation{IBM Research Europe - Zurich, 8803 Ruschlikon, Switzerland}
\author{Vicky Hasse}\affiliation{ Max Planck Institute for Chemical Physics of Solids, 01187 Dresden, Germany}
\author{Claudia Felser}\affiliation{ Max Planck Institute for Chemical Physics of Solids, 01187 Dresden, Germany}
\author{Heinz Schmid}\affiliation{IBM Research Europe - Zurich, 8803 Ruschlikon, Switzerland}
\author{Cezar Zota}\affiliation{IBM Research Europe - Zurich, 8803 Ruschlikon, Switzerland}
\author{Bernd Gotsmann} \email{bgo@zurich.ibm.com} \affiliation{IBM Research Europe - Zurich, 8803 Ruschlikon, Switzerland}

\date{\today}
\maketitle

\textbf{
The scaling of Si transistor technology has resulted in a remarkable improvement in the performance of integrated circuits over the last decades. However, scaled transistors also require reduced electrical interconnect dimensions, which lead to greater losses and power dissipation at circuit level. This is mainly caused by enhanced surface scattering of charge carriers in copper interconnect wires at dimensions below 30 nm. A promising approach to mitigate this issue is to use directional conductors, i.e. materials with anisotropic Fermi surface, where proper alignment of crystalline orientation and transport direction can minimize surface scattering. In this work, we perform a resistivity scaling study of the anisotropic semimetal NbP as a function of crystalline orientation. We use here focused ion beam to pattern and scale down NbP crystallites to dimensions comparable to the electron scattering length at cryogenic temperatures. The experimental transport properties are correlated with the Fermi surface characteristics through a theoretical model, thus identifying the physical mechanisms that influence the resistivity scaling of anisotropic conductors. Our methodology provides an effective approach for early evaluation of anisotropic materials as future ultra-scalable interconnects, even when they are unavailable as epitaxial films.
}

\section{Introduction}
Current estimates indicate that information and communication technology (ICT) accounts for approximately 8–10\% of global electricity consumption \cite{gelenbe_electricity_2023}, with this share expected to rise significantly in the coming decades \cite{andrae_global_2015, andrae_projecting_2019, andrae_new_2020}. Among the primary contributors are Si CMOS integrated circuits, whose energy requirements are projected to surpass global energy production by 2040 \cite{dcadmin_rebooting_2015}, highlighting the urgent need for innovative technologies for energy-efficient ICT systems. One major cause of dissipation in CMOS chips are the copper interconnects, which connect billions of transistors and deliver power and signals to them. As transistors have approached nanometer sizes, following Moore's law, so have interconnects as well, but while transistor performance improves with miniaturization, interconnect performance instead degrades. This is caused by an increased resistance due to a reduced cross-sectional area and enhanced diffusive electron scattering at the surfaces, resulting in higher power consumption and signal delay \cite{gall_materials_2021}.

The shrinking dimensions of interconnects pose a significant challenge to the continued use of copper, as its performance deteriorates due to increased diffusive scattering at interfaces and grain boundaries \cite{gall_materials_2021}. To address these issues, significant research efforts have explored strategies to reduce interface scattering \cite{van_der_veen_barrierliner_2016, chawla_specular_2009, zheng_ni_2014, zheng_electron_2016, gall_materials_2021}, and improve grain boundary properties \cite{barmak_surface_2014, cesar_reducing_2016, cesar_calculated_2014, kim_large_2010, rickman_simulation_2013}. In addition, alternative materials that could show lower resistivity at dimensions below 30 nm are being investigated \cite{gall_search_2020, dutta_ruthenium_2017, choi_electron_2012, kamineni_tungsten_2016, chawla_resistance_2016, zhang_ruthenium_2016, kelly_experimental_2016, wen_ruthenium_2016, dutta_thickness_2017, zhang_methods_2017, hu_electromigration_2017}, including metals with reduced scattering lengths \cite{gall_materials_2021, gall_electron_2016, gall_search_2020, gall_metals_2018}, topological properties \cite{gall_materials_2021, han_1d_2021, zhang_ultrahigh_2019, lien_unconventional_2023, rocchino_unconventional_2024} and anisotropic conductivity \cite{kumar_ultralow_2022, zheng_anisotropic_2017, hashimoto_anisotropy_1998}.

\begin{figure*}
\includegraphics[width=\linewidth]{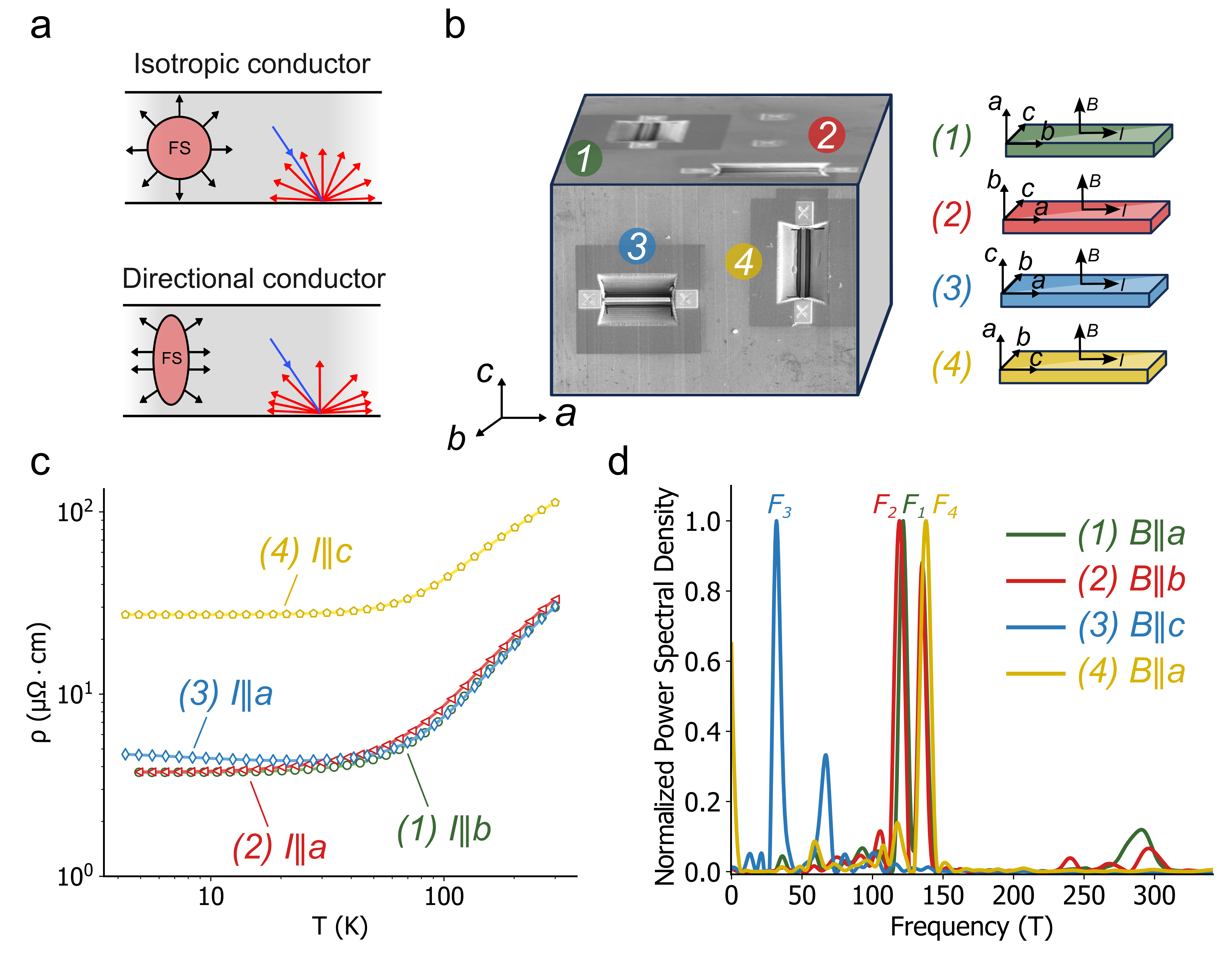}
\caption{\label{fig:panel1} \bf{Anisotropic transport in NbP.} \textbf{(a)} In isotropic conductors (top), a spherical Fermi surface (FS, as labeled in the figure) leads to uniform electron velocity distribution, resulting in random velocity orientations after surface scattering. In directional conductors (bottom), an anisotropic oriented FS aligns most electron velocities close to the transport direction, increasing the mean free path between surface scattering events. \textbf{(b)} Scanning Electron Microscope (SEM) images schematically showing the relative orientation of four NbP lamellae extracted from a single crystals. The chosen axes orientations enable transport measurements with the current and magnetic field applied along any principal axis. \textbf{(c)} Temperature-dependent electrical resistivity of the NbP samples, with transport directions aligned along different principal axes of the NbP crystal. \textbf{(d)} Power spectral density (normalized to largest peak) of SdH oscillations extracted from magnetoresistance measurements on the NbP samples, with varying directions of the applied magnetic field, \(\mathbf{B}\).}
\end{figure*}

Materials with highly anisotropic conductivity have recently been predicted to outperform copper in square wires narrower than 25 nm \cite{kumar_ultralow_2022}. This advantage arises from leveraging Fermi surface anisotropy by orienting the material so that most states on the Fermi surface have velocities aligned with the transport direction while minimizing those with velocities orthogonal to the conductor’s surfaces, as shown in Figure \ref{fig:panel1}a. Consequently, surface scattering has a reduced impact, resulting in a resistivity that increases slowly even at nanometer-scale widths, unlike in isotropic conductors such as copper.

In this work, we investigate the link between anisotropic Fermi surfaces and orientation-dependent resistivity scaling using the anisotropic semimetal NbP, and provide a straightforward method for evaluating the performance of materials in this context. We employ focused ion beam (FIB) to fabricate and thin down samples with different crystallographic orientations, all extracted from the same macroscopic single crystal. Leveraging the long scattering length of high-quality single crystals, surface scattering effects can be investigated in the micrometer-sized samples fabricated using FIB. This approach eliminates the need for epitaxial thin film growth, which may not be available for the target material, and allows for easy alignment of the crystalline directions as required by the transport geometry. Moreover, by measuring the Shubnikov-de Haas (SdH) effect, feasible in high-quality single crystals, we determine Fermi surface characteristics \cite{schoenberg_magnetic_2009}, and correlate those with resistivity scaling properties. This was not possible in a related study on tungsten \cite{zheng_anisotropic_2017}, due to lower crystal quality and a reduced scattering length in thin films. Our results show that aligning anisotropic conductors along preferential crystalline axes mitigates resistivity increases in down-scaled metallic wires at low temperatures, where transport enters the quasi-ballistic regime. We develop a model linking Fermi surface and velocity anisotropies to resistivity scaling, providing insights into the behavior of scaled anisotropic conductors. This model offers a rapid and accessible approach to evaluate anisotropic materials for low-dissipation interconnect applications through easily measurable quantities.

\section{\label{sec:results}Results}
\subsection{\label{sec:direction}Directional Dependent Transport}
To characterize the anisotropic transport properties of NbP, four lamellae were extracted from two facets of a bulk single crystal, as shown in figure \ref{fig:panel1}b, and shaped into Hall bars on the micrometer scale using FIB. The orientations were chosen to enable transport measurements with the current or the out-of-plane magnetic field applied along any principal axis. The Hall bar dimensions are reported in the Supplementary Information. FIB processing facilitates directional-dependent measurements by allowing for the fabrication of micro-Hall bars along various crystalline directions. In contrast, such measurements are particularly challenging for thin films, where measuring out-of-plane resistivity is difficult, and for single crystals grown via chemical vapour transport (CVT), which often do not permit electrical contacts on each high-symmetry facet. Measurements performed on these relatively large devices also provide a reliable approximation of bulk properties.

To establish the high- and low-conductivity axes of NbP, resistivity measurements as a function of temperature were performed on the four Hall bars. The results, shown in figure \ref{fig:panel1}c, reveal that the \(a\) and \(b\) axes exhibit nearly identical resistivity, while the \(c\) axis consistently shows an order of magnitude higher resistivity across all temperatures. Minor deviations between the \(a\) and \(b\) axes at low temperatures are likely attributable to sample preparation and surface effects influencing resistivity. These findings indicate that the \(a\) and \(b\) axes are effectively equivalent in terms of resistivity.

To experimentally characterize the Fermi surface of NbP, SdH oscillations were isolated from magnetoresistance measurements performed on Hall bars by varying the magnetic field applied along the out-of-plane axis. The associated measured power spectral densities, shown in figure \ref{fig:panel1}d, reveal a lower oscillation frequency when the field is applied along the \(c\)-axis. Since the oscillation frequency is directly proportional to the area, perpendicular to the applied magnetic field, enclosed by an extremal orbit on the Fermi surface \cite{onsager_interpretation_1952, roth_semiclassical_1966, ashcroft_solid_1976}, this indicates that the Fermi surface of NbP is elongated along the \(c\)-axis. This finding is consistent with previous literature \cite{schindler_effect_2020, shekhar_extremely_2015, klotz_quantum_2016, lee_fermi_2015, balduini_origin_2024} and with the resistivity measurements in figure \ref{fig:panel1}c. This is because an elongated Fermi surface along \(c\) implies fewer states with velocity in this direction, leading to higher resistivity. Such anisotropy in the Fermi surface makes NbP a suitable candidate for testing the directional conductor strategy, where an anisotropic Fermi surface is essential.

\begin{figure}
\includegraphics[width=\linewidth]{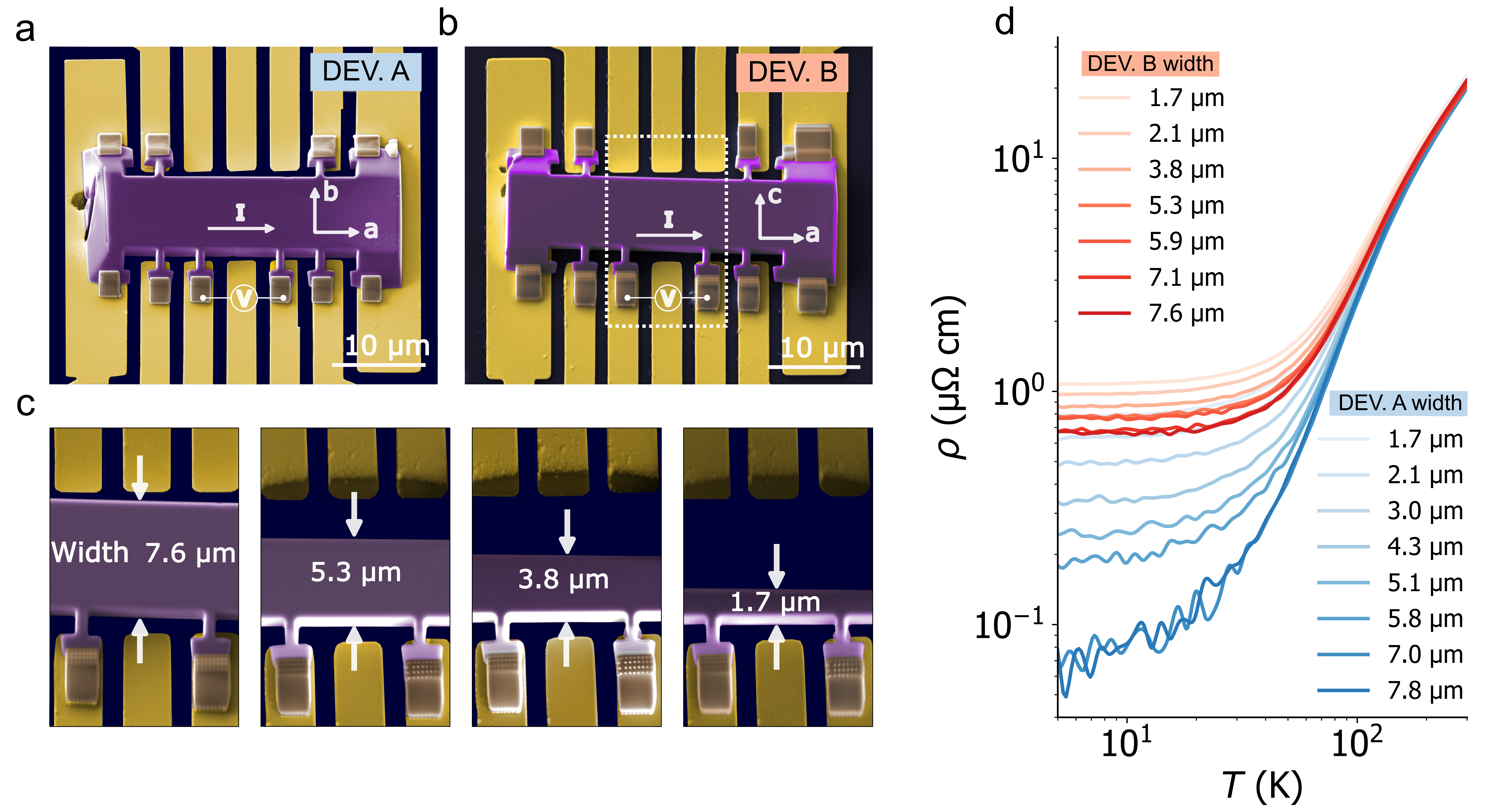}
\caption{\label{fig:panel2} \bf{Focused ion beam scaling in NbP}. \textbf{(a)}-\textbf{(b)} False-colored SEM images of the NbP samples (purple) positioned on the pre-patterned gold electrodes (gold) on \(\mathbf{Si}\)/\(\mathbf{SiO_2}\) substrates (blue), with Pt-contacts (brown) deposited using ion beam deposition to create electrical contacts. The crystalline orientation and the configuration of the electrical contacts for the two devices are shown. \textbf{(c)} Focused Ion Beam (FIB) was used to systematically reduce the width of the two devices to study the dependence of resistivity on sample size. Representative SEM images of the scaling process for device B are shown here. \textbf{(d)} Resistivity as a function of temperature for different widths in the two devices.}
\end{figure}

\subsection{\label{sec:scaling}Orientation and Size Dependence of Longitudinal Resistivity}
To investigate the influence of sample dimensions on resistivity and how it depends on crystal orientation, we measure the resistivity of two devices while progressively reducing their width. For a directional conductor, the aim is to align low-conductivity axes orthogonal to the transport direction to minimize electron surface scattering and hence mitigate the resistivity increase as the sample dimensions are reduced. To test this, two additional Hall bars, labeled A and B, were fabricated from the bulk crystal, both with the conduction axis aligned along the high-conductivity axis \(a\). In contrast, the axes aligned with the width differ between the devices: device A has the high-conductivity axis \(b\) along this direction (figure \ref{fig:panel2}a), while device B has the low-conductivity axis \(c\) (figure \ref{fig:panel2}b). These orientations correspond to those of devices 3 and 2 in figure \ref{fig:panel1}, respectively. The widths of the two devices were progressively reduced, as illustrated in figure \ref{fig:panel2}c, and resistivity measurements were taken at each step, with data reported in figure \ref{fig:panel2}d. The exact dimensions of the devices are provided in the Supplementary Information.

The resistivities of the two devices are nearly identical and width-independent at high temperatures, but increase with decreasing width at low temperatures. This behavior arises because, at cryogenic temperatures, the scattering length approaches the device size, making surface effects significant. For the same single crystal, a previous study estimated the scattering length to be approximately 5 µm below 20 K \cite{balduini_probing_2024}, highlighting the regime where surface scattering becomes relevant. Device B consistently exhibits higher resistivity at low temperatures for all widths, which we attribute to its smaller thickness (\(h_A = 3.25\) \textmu m, \(h_B = 2.20\) \textmu m), leading to increased surface scattering along the out-of-plane direction. Nevertheless, the resistivity increase with decreasing width is more pronounced in device A than in device B, as expected, since device B has the low-conductivity axis \(c\) aligned with the width.

\section{Theoretical framework and data analysis}

\subsection{\label{sec:rho_model}Resistivity Model for Directional Conductors}
To explore the impact of device size variations and the orientation of an anisotropic Fermi surface relative to material surfaces on resistivity, we derive a resistivity model for directional conductors. The model is based on the Boltzmann transport equation and considers thin wires with a rectangular cross-section, as described in \cite{kumar_ultralow_2022}. We assume surface scattering to be purely diffusive, consistent with the amorphous layers induced by FIB microstructuring \cite{bachmann_inducing_2017}, which has been shown to promote momentum non-conserving scattering at material surfaces \cite{balduini_probing_2024}.

A readily interpretable expression for the size-dependent resistivity is obtained by considering the large-device approximation. Although this approximation is strictly valid only for devices much larger than the scattering length, prior studies suggest it remains accurate for smaller sizes, down to approximately one-tenth of the scattering length \cite{kumar_ultralow_2022}. As detailed in the Supplementary Information, the resistivity in this limit simplifies to:
\begin{equation}\label{eq:res_approx}
    \rho\left(w,h\right)\approx\rho_0\left[1+\frac{g_2(\hat{j},\hat{n}_1)}{g_1(\hat{j})w}+\frac{g_2(\hat{j},\hat{n}_2)}{g_1(\hat{j})h}\right]
\end{equation}
Here, \(\hat{j}\) is the unit vector along the transport direction, \(\hat{n}_1\) and \(\hat{n}_2\) are the unit vectors along the width \(w\) and thickness \(h\) of the wire, respectively, and \(\rho_0\) is the bulk resistivity. The material and orientation specific properties are captured by the functions \(g_1(\hat{j})\) and \(g_2(\hat{j},\hat{n})\), defined as:

\begin{subequations}\label{eq:gs}
    \begin{equation}\label{eq:g1}
        g_1(\hat{j})=e^2\sum_b\int_{BZ}\frac{g_sd^3k}{\left(2\pi\right)^3}\left(-f'_0\left(E_{\mathbf{k}b}\right)\right)\left(v_{\mathbf{k}b}^j\right)^2
    \end{equation}
    \begin{equation}\label{eq:g2}
        g_2(\hat{j},\hat{n})=e^2\sum_b\int_{BZ}\frac{g_sd^3k}{\left(2\pi\right)^3}\left(-f'_0\left(E_{\mathbf{k}b}\right)\right)\left(v_{\mathbf{k}b}^j\right)^2\left|v_{\mathbf{k}b}^n\right|\tau_{\mathbf{k}b}
    \end{equation}
\end{subequations}
Here \(e\) is the electron charge, \(g_s=2\) is the spin degeneracy factor, while \(E_{\mathbf{k}b}\), \(v_{\mathbf{k}b}\) and \(\tau_{\mathbf{k}b}\) represent the electronic energies, velocities, and scattering times for band \(b\) and wavevector \(\mathbf{k}\). The derivative of the Fermi-Dirac distribution \(f_0'(E_{\mathbf{k}b})\) confines the momentum-space integrals (\(d^3k\)) over the Brillouin zone (BZ) to the neighborhood of the Fermi surface, as expected in transport phenomena.

To mitigate resistivity increases with decreasing wire diameter, the material should be aligned to minimize \(\left|v_{\mathbf{k}b}^n\right|\), the velocity component perpendicular to the transport direction, thus reducing \(g_2(\hat{j},\hat{n})\). This is the fundamental idea behind directional conductors, which reduce surface scattering by directing more electrons along the transport axis and away from the device surfaces, thus limiting resistivity growth.

\subsection{\label{sec:FS_model}Ellipsoidal Fermi Surface Model}
To compute the velocities required for equations (\ref{eq:g1}) and (\ref{eq:g2}), we adopt an anisotropic linear band model. This approach reflects the linear band dispersion near the Weyl nodes, which are characteristic of NbP as a Weyl semimetal, and captures its anisotropic transport properties and Fermi surface geometry. The band dispersion in this model is expressed as:

\begin{equation} 
    E=\hbar v_F\sqrt{k_a^2+k_b^2+\frac{k_c^2}{\gamma^2}}
\end{equation}

Here \(E\) is the energy, \(\hbar\) is the reduced Planck constant, \(k_i\) are the components of the \(\mathbf{k}\)-vector in reciprocal space, \(v_F\) is a constant with units of velocity, and \( \gamma \) quantifies the anisotropy along the \( c \)-direction.

The constant energy surfaces form prolate ellipsoids elongated along the \(c\)-axis for \(\gamma > 1\), representing the simplest deviation from spherical isotropic Fermi surfaces (where \(\gamma = 1\)). This shape is supported by experimental data from the first section, where it emerged that the \(c\)-axis had both lower electrical conductivity and lower frequency of SdH oscillations. Ellipsoidal surfaces have also been shown to approximate the behavior of anisotropic conductors effectively \cite{fuchser_anisotropic_1971, mackey_electron_1968}, and their impact on thin-film resistivity have been explored theoretically \cite{price_anisotropic_1960, parrott_new_1965, ham_electrical_1960, englman_electrical_1956}. In this work, we extend the model to the case of rectangular thin-wires.

The electron velocity is obtained as the gradient of the energy dispersion. Details of the derivation, conducted in prolate spheroidal coordinates, are provided in the Supplementary Information. In this coordinate system, the energy depends only on one coordinate, while the other two sweep surfaces of constant energy. The resulting expression for the velocity is:
\begin{equation}\label{eq:velocities}
    \mathbf{v}=v_F\begin{pmatrix}
        \sin\left(\theta\right)\cos\left(\varphi\right)\\
        \sin\left(\theta\right)\sin\left(\varphi\right)\\
        \frac{\cos\left(\theta\right)}{\gamma}
    \end{pmatrix}
\end{equation}
Equation (\ref{eq:velocities}) clarifies the role of the model parameters \(v_F\) and \(\gamma\): \(v_F\) is the Fermi velocity along the crystallographic axes \(a\) and \(b\), while \(v_F/\gamma\) is the velocity along the \(c\) axis, accounting for the anisotropic nature of the Fermi surface.

To determine the resistivity dependence on device size for configurations A and B, we substitute equation (\ref{eq:velocities}) into equations (\ref{eq:g1}) and (\ref{eq:g2}). For device A, the orientation vectors are defined as \(\hat{n}_1\) along \(b\) and \(\hat{n}_2\) along \(c\), while for device B, these orientations are reversed. In both cases, the transport direction vector \(\hat{j}\) aligns with the \(a\) axis. With this setup, we can calculate \(g_1(\hat{j})\), \(g_2(\hat{j},\hat{n}_1)\), and \(g_2(\hat{j},\hat{n}_2)\) explicitly, using a few additional assumptions.

The scattering time, \(\tau_{\mathbf{k}b}\), is assumed to be momentum-independent to enable analytical tractability: \(\tau_{\mathbf{k}b} = \tau\). This simplification is consistent with prior studies, where it was shown to produce results comparable to alternative approximations, such as treating the scattering length \(l_e\) as momentum-independent \cite{zheng_anisotropic_2017}.

A second assumption concerns the Fermi-Dirac distribution at \(0 \, \text{K}\), where its first derivative is approximated as a delta function centered at the Fermi energy, effectively restricting the integration domain to the Fermi surface. With these assumptions, the resulting expressions for devices A and B are:
\begin{subequations}\label{eq:for_fits}
    \begin{equation}\label{eq:for_fits_a}
    \rho_A(w,h) \approx \rho_0 \left[ 1 + \frac{3v_F\tau}{8w} + \frac{3v_F\tau}{8h}\frac{1}{\gamma} \right]=\rho_0 \left[ 1 + \frac{3l_{e_b}}{8w} + \frac{3l_{e_c}}{8h} \right]
    \end{equation}
    \begin{equation}\label{eq:for_fits_b}
    \rho_B(w,h) \approx \rho_0 \left[ 1 + \frac{3v_F\tau}{8w}\frac{1}{\gamma} + \frac{3v_F\tau}{8h} \right]=\rho_0 \left[ 1 + \frac{3l_{e_c}}{8w} + \frac{3l_{e_b}}{8h} \right]
    \end{equation}
\end{subequations}

Here, the scattering length \(l_{e_i}\) along the \(i\) axis is calculated as the product of the velocity \(v_i\) along the \(i\) axis with the isotropic scattering time \(\tau\), such that \(l_{e_i}=v_i\tau\). Using equation (\ref{eq:velocities}), where \(v_c=v_b/\gamma\), it follows that \(l_{e_c}=l_{e_b}/\gamma=v_F\tau/\gamma\). Derivations of these formulas, along with a discussion of the effects of removing the \(0 \, \text{K}\) assumption on the Fermi-Dirac distribution, are provided in the Supplementary Information.

Equations (\ref{eq:for_fits_a}) and (\ref{eq:for_fits_b}) reveal that the resistivity scaling with size reduction is directly proportional to the scattering lengths \(l_{e_i}\) along the respective directions. This observation aligns with earlier theoretical findings \cite{ham_electrical_1960}. Furthermore, since \(l_{e_c}=l_{e_b}/\gamma\), the resistivity increase due to width reduction is mitigated in device B by a factor of \(\gamma\) compared to device A, which reflects the anisotropy \(\gamma\) in the Fermi surface.

\begin{figure*}
\includegraphics[width=\linewidth]{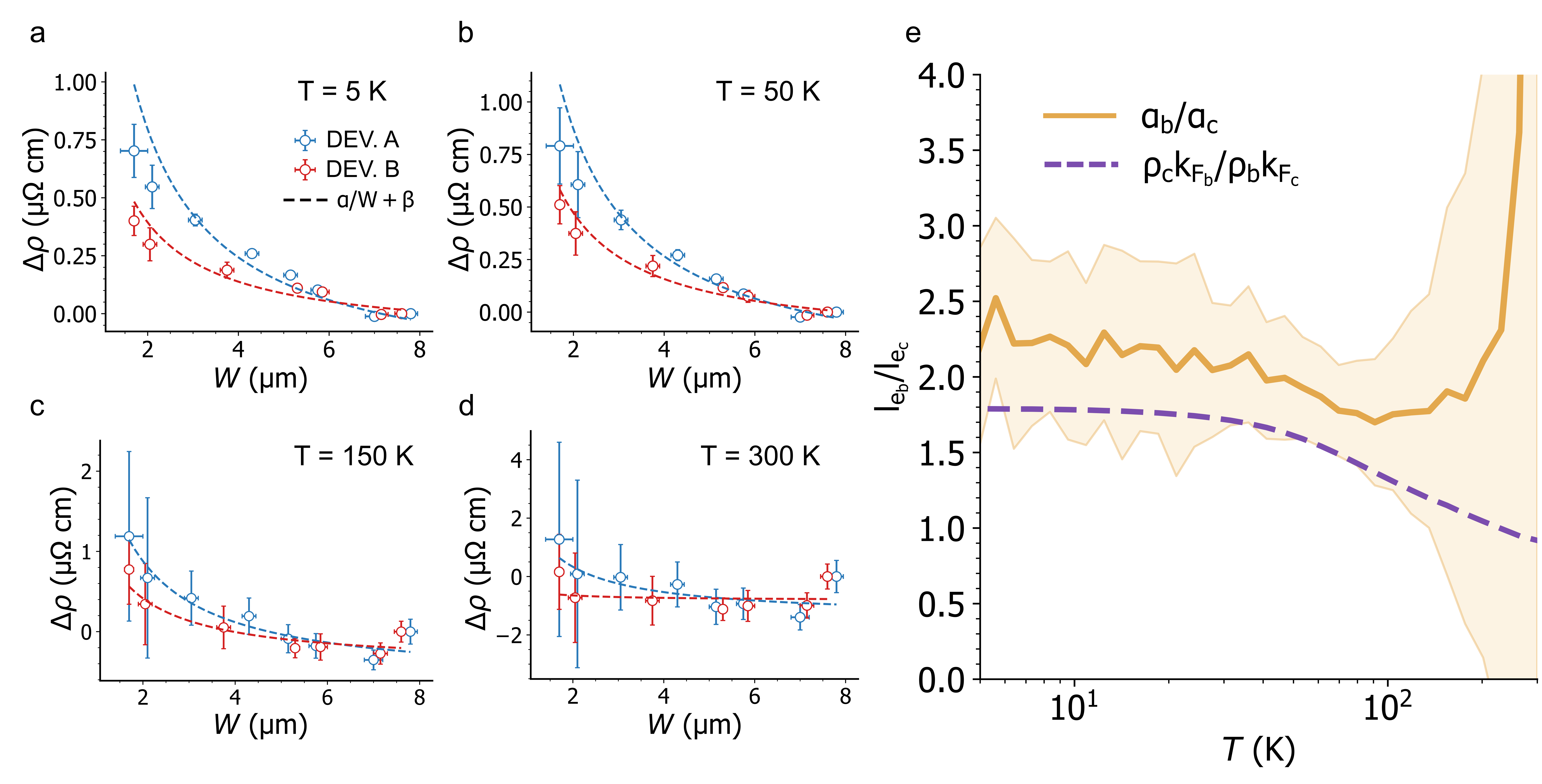}
\caption{\label{fig:Panel_3} \bf{Resistivity scaling in NbP} \textbf{(a)}-\textbf{(d)} Resistivity increase \(\mathbf{\Delta\rho}\) as a function of width \(\mathbf{W}\) for two devices at four different temperatures. \(\mathbf{\Delta\rho}\) is calculated relative to the resistivity at the largest width. The best-fit curves for both datasets, derived from equations (\ref{eq:for_fits_a}) and (\ref{eq:for_fits_b}), are also shown. \textbf{(e)} Scattering length ratio along the \(b\) and \(c\) directions of NbP, providing a measure of its resistivity scaling anisotropy. The ratio is shown as determined from two methods: resistivity scaling coefficients (gold, with shaded areas indicating uncertainties) and an estimate using resistivity and SdH data within Drude’s model (purple).}
\end{figure*}

\subsection{\label{sec:exp_comp}Experimental Comparison}
An experimental value for \(\gamma=l_{e_b}/l_{e_c}\) is determined by fitting the measured resistivity data as function of the width using equations (\ref{eq:for_fits_a}) and (\ref{eq:for_fits_b}), as shown in Fig. \ref{fig:Panel_3}a. Since the thickness \(h\) was not varied in the experiments, the terms \(3l_{e_c}/8h\) in equation (\ref{eq:for_fits_a}) and \(3l_{e_b}/8h\) in equation (\ref{eq:for_fits_b}) simplify to constants. The resulting equation for fitting the experimental data is given by $\rho_i(w)=\alpha_i/w+\beta_i$, where  \(\alpha_i\) and \(\beta_i\) are the fitting parameters.
The fits are shown in figure \ref{fig:Panel_3}a-d for various temperatures. As expected, device A exhibits a larger resistivity increase compared to device B. Additionally, no significant scaling is observed at room temperature, as electrons effectively become insensitive to the device surfaces due to the reduced scattering length compared to the device cross section. To determine the scaling anisotropy, which equals the scattering length ratio $\gamma=l_{e_b}/l_{e_c}$, we use the ratio of the coefficients \(\alpha_i\), as shown in equations (\ref{eq:for_fits_a}) and (\ref{eq:for_fits_b}): $\gamma=l_{e_b}/l_{e_c}=\alpha_A/\alpha_B$. The obtained values of \(l_{e_b}/l_{e_c}\)  are shown in figure \ref{fig:Panel_3}e.

\subsection{Comparison with Resistivity and Fermi Momentum Anisotropy}
We now outline a method to correlate the scaling results with measurements of resistivity and Fermi surface anisotropy. Using the definition of bulk scattering length \(l_e\), we derive:
\begin{equation}
    l_{e_i}=v_{F_i}\tau=\frac{p_{F_i}\tau}{m_i}=\frac{\hbar k_{F_i}\tau}{m_i}=\frac{\hbar k_{F_i}}{ne^2}\frac{ne^2\tau}{m_i}=\frac{\hbar k_{F_i}}{ne^2}\frac{1}{\rho_i}
\end{equation}
The final step in this derivation employs Drude’s formula \cite{ashcroft_solid_1976}. Here \(v_{F_i}\), \(p_{F_i}\), \(k_{F_i}\) are the \(i\)-th component of the Fermi velocity, momentum and \(\mathbf{k}\)-vector, \(\tau\) is the scattering time, \(m_i\) is the effective mass along direction \(i\), \(n\) is the carrier density and \(\rho_i\) is the bulk resistivity along direction \(i\). This derivation allows us to relate the ratio of scattering lengths, estimated from the resistivity versus length scaling experiment, to the following expression:
\begin{equation}
    \gamma=\frac{l_{e_b}}{l_{e_c}}=\frac{\rho_c k_{F_b}}{\rho_b k_{F_c}}
\end{equation}
The ratio \(\rho_c/\rho_b\) is determined from the temperature-dependent resistivity measurements on the four Hall bars, while the ratio of k-components \(k_{F_b} / k_{F_c}\) is calculated as the temperature-independent ratio of the Fermi surface areas derived from the frequency of the SdH oscillations, under the ellipsoidal Fermi surface approximation. This second estimate of the scattering length ratio is plotted alongside the results from scaling experiments in figure \ref{fig:Panel_3}e, bringing together all experimental data from this study in a single graph. Both methods show a scattering length ratio of approximately 2 over most of the temperature range, demonstrating overall consistency. At higher temperatures, when transport enters 
the diffusive regime, the scaling method fails to provide reliable estimates with low uncertainties.

The strong agreement between the two methods demonstrates that the anisotropy in the resistivity scaling coefficients quantitatively reflects the Fermi surface anisotropy, thus confirming the potential of anisotropic conductors in mitigating resistivity increases in interconnects. Moreover, the comparison in figure \ref{fig:Panel_3}e highlights a straightforward and practical approach for estimating material scaling anisotropy by combining resistivity measurements and estimates of the wave vector \(\mathbf{k_F}\).

\section{Conclusions}
We have in this work demonstrated that materials with anisotropic Fermi surfaces can be leveraged to realize directional conductors, where the resistivity increase due to surface scattering is mitigated. The anisotropic semimetal NbP was studied in this context, given its approximately elliptical Fermi surface. The mesoscopic NbP crystal that was scaled down along the \(c\)-axis showed a substantially smaller increase of resistivity than the sample scaled down along the \(b\)-axis. Through a Fermi surface-sensitive resistivity model, we established that the critical anisotropic scaling parameter is the ratio of scattering lengths along different crystallographic directions. This ratio can also be expressed in terms of a ratio between the product of Fermi k-vectors and resistivities along different directions, which provides the link between conductivity and Fermi surface geometry. Our study on NbP mesoscopic crystals confirmed this relationship, matching an experimentally determined scattering length ratio of 2 with the calculated anisotropy of the ellipsoidal NbP Fermi surface. This work offers experimental proof of the promising scaling properties of anisotropic conductors for electrical interconnect applications, and also provides a straightforward framework for assessing the electrical performance of these materials.

\section{\label{sec:Methods}Methods}

\subsection{\label{sec:growth}Crystal Growth}
High-quality bulk single crystals of NbP were grown at MPI CPfS using a chemical vapor transport reaction with iodine as the transport agent. A polycrystalline NbP powder was synthesized by directly reacting niobium (Chempur 99.9\%) with red phosphorus (Heraeus 99.999\%) within an evacuated fused silica tube for \num{48} \unit{h} at \num{800}\unit{\degreeCelsius}. The growth of bulk single crystals of NbP was then initialized from this powder by chemical vapor transport in a temperature gradient, starting from \num{850}\unit{\degreeCelsius} (source) to \num{950}\unit{\degreeCelsius} (sink) and a transport agent with a concentration of \num{13.5} \unit{mg.cm^{-3}} iodine (Alfa Aesar 99.998\%).

\subsection{\label{sec:fabrication}Sample Fabrication}
Microscopic bars were prepared from a NbP single crystal using focused ion beam (FIB) microstructuring \cite{moll_focused_2018}. This method was used to achieve high aspect-ratio samples with desired geometry, crystalline orientation, and uniform magnetic field distribution along the sample. Although the process alters a thin surface layer \cite{bachmann_inducing_2017}, bulk properties remain unaffected, as demonstrated by previous observations \cite{balduini_probing_2024}. After milling, the samples were placed on a silicon chip with a silicon oxide spacer and patterned gold lines. Platinum contacts were deposited using ion-assisted chemical vapor deposition, resulting in contact resistances of approximately \num{15} \unit{\ohm}. FIB was further used to systematically narrow the Hall bars, allowing for the investigation of transport property changes as a function of sample size.

\subsection{\label{sec:tansport}Electrical Transport Measurements}
Electrical measurements were conducted in a cryostat (Dynacool from Quantum Design) using external lock-in amplifiers (MFLI from Zurich Instruments). An AC current with a constant amplitude of approximately \num{100} \unit{\micro A} at \num{211} \unit{Hz} was applied for both the scaling and resistivity/SdH experiments.

\subsection{\label{sec:SdH}Shubnikov--de Haas Oscillations Analysis}
Shubnikov--de Haas oscillations have been isolated by performing the second derivative of the magnetoresistive data, from 2 T to 9 T. The power spectral density has been found by performing a fast Fourier transform on the oscillations plotted as \(1/B\).

\section{Acknowledgements}

We are grateful to T. Paul and J. Schadt for insightful discussions. This work was supported by the Horizon Europe project AttoSwitch, the Swiss State Secretariat for Education, Research and Innovation (SERI) under contract number 23.00594. F.B. and B.G. acknowledge the SNSF project HYDRONICS under the Sinergia grant (No.189924). C. Z. and L. R. acknowledge the SNSF Ambizione programme (no. 193636) and the SNSF grant DiracSource, (no. IZKSZ2\_218591). We wish to acknowledge the support of the Cleanroom Operations Team of the Binning and Rohrer Nanotechnology Center (BRNC).

\section{Author Contribution}
F.B. conceived the experiment. G.M. and F.B. developed the theory. C.F., V.H. grew the crystals. F.B. fabricated the samples. G.M. and F.B scaled the sample using focused ion beam. G.M. and F.B. performed the measurements and data analysis. G.M., F.B., N.D., L.R., C.Z., H.S., and B.G. interpreted the data. G.M. and F.B. wrote the manuscript with inputs from all authors. 

\section{COMPETING INTERESTS}
The authors declare no competing interests.


\begin{thebibliography}{53}%
\makeatletter
\providecommand \@ifxundefined [1]{%
 \@ifx{#1\undefined}
}%
\providecommand \@ifnum [1]{%
 \ifnum #1\expandafter \@firstoftwo
 \else \expandafter \@secondoftwo
 \fi
}%
\providecommand \@ifx [1]{%
 \ifx #1\expandafter \@firstoftwo
 \else \expandafter \@secondoftwo
 \fi
}%
\providecommand \natexlab [1]{#1}%
\providecommand \enquote  [1]{``#1''}%
\providecommand \bibnamefont  [1]{#1}%
\providecommand \bibfnamefont [1]{#1}%
\providecommand \citenamefont [1]{#1}%
\providecommand \href@noop [0]{\@secondoftwo}%
\providecommand \href [0]{\begingroup \@sanitize@url \@href}%
\providecommand \@href[1]{\@@startlink{#1}\@@href}%
\providecommand \@@href[1]{\endgroup#1\@@endlink}%
\providecommand \@sanitize@url [0]{\catcode `\\12\catcode `\$12\catcode `\&12\catcode `\#12\catcode `\^12\catcode `\_12\catcode `\%12\relax}%
\providecommand \@@startlink[1]{}%
\providecommand \@@endlink[0]{}%
\providecommand \url  [0]{\begingroup\@sanitize@url \@url }%
\providecommand \@url [1]{\endgroup\@href {#1}{\urlprefix }}%
\providecommand \urlprefix  [0]{URL }%
\providecommand \Eprint [0]{\href }%
\providecommand \doibase [0]{https://doi.org/}%
\providecommand \selectlanguage [0]{\@gobble}%
\providecommand \bibinfo  [0]{\@secondoftwo}%
\providecommand \bibfield  [0]{\@secondoftwo}%
\providecommand \translation [1]{[#1]}%
\providecommand \BibitemOpen [0]{}%
\providecommand \bibitemStop [0]{}%
\providecommand \bibitemNoStop [0]{.\EOS\space}%
\providecommand \EOS [0]{\spacefactor3000\relax}%
\providecommand \BibitemShut  [1]{\csname bibitem#1\endcsname}%
\let\auto@bib@innerbib\@empty
\bibitem [{\citenamefont {Gelenbe}(2023)}]{gelenbe_electricity_2023}%
  \BibitemOpen
  \bibfield  {author} {\bibinfo {author} {\bibfnamefont {E.}~\bibnamefont {Gelenbe}},\ }\href {https://doi.org/10.1145/3613207} {\bibfield  {journal} {\bibinfo  {journal} {Ubiquity}\ }\textbf {\bibinfo {volume} {2023}},\ \bibinfo {pages} {1:1} (\bibinfo {year} {2023})}\BibitemShut {NoStop}%
\bibitem [{\citenamefont {Andrae}\ and\ \citenamefont {Edler}(2015)}]{andrae_global_2015}%
  \BibitemOpen
  \bibfield  {author} {\bibinfo {author} {\bibfnamefont {A.}~\bibnamefont {Andrae}}\ and\ \bibinfo {author} {\bibfnamefont {T.}~\bibnamefont {Edler}},\ }\href {https://doi.org/10.3390/challe6010117} {\bibfield  {journal} {\bibinfo  {journal} {Challenges}\ }\textbf {\bibinfo {volume} {6}},\ \bibinfo {pages} {117} (\bibinfo {year} {2015})}\BibitemShut {NoStop}%
\bibitem [{\citenamefont {Andrae}(2019)}]{andrae_projecting_2019}%
  \BibitemOpen
  \bibfield  {author} {\bibinfo {author} {\bibfnamefont {A.~S.~G.}\ \bibnamefont {Andrae}},\ }\href {https://doi.org/10.13140/RG.2.2.25103.02724} {{\selectlanguage {English}\bibinfo {title} {Projecting the chiaroscuro of the electricity use of communication and computing from 2018 to 2030}}} (\bibinfo {year} {2019}),\ \bibinfo {note} {publisher: Unpublished}\BibitemShut {NoStop}%
\bibitem [{\citenamefont {Andrae}(2020)}]{andrae_new_2020}%
  \BibitemOpen
  \bibfield  {author} {\bibinfo {author} {\bibfnamefont {A.~S.~G.}\ \bibnamefont {Andrae}},\ }\href {https://doi.org/10.30538/psrp-easl2020.0038} {\bibfield  {journal} {\bibinfo  {journal} {Engineering and Applied Science Letters}\ }\textbf {\bibinfo {volume} {3}},\ \bibinfo {pages} {19} (\bibinfo {year} {2020})}\BibitemShut {NoStop}%
\bibitem [{\citenamefont {dcadmin}(2015)}]{dcadmin_rebooting_2015}%
  \BibitemOpen
  \bibfield  {author} {\bibinfo {author} {\bibnamefont {dcadmin}},\ }\href {https://www.semiconductors.org/resources/rebooting-the-it-revolution-a-call-to-action-2/} {{\selectlanguage {English}\bibinfo {title} {Rebooting the {IT} {Revolution}: {A} {Call} to {Action}}}} (\bibinfo {year} {2015})\BibitemShut {NoStop}%
\bibitem [{\citenamefont {Gall}\ \emph {et~al.}(2021)\citenamefont {Gall}, \citenamefont {Cha}, \citenamefont {Chen}, \citenamefont {Han}, \citenamefont {Hinkle}, \citenamefont {Robinson}, \citenamefont {Sundararaman},\ and\ \citenamefont {Torsi}}]{gall_materials_2021}%
  \BibitemOpen
  \bibfield  {author} {\bibinfo {author} {\bibfnamefont {D.}~\bibnamefont {Gall}}, \bibinfo {author} {\bibfnamefont {J.~J.}\ \bibnamefont {Cha}}, \bibinfo {author} {\bibfnamefont {Z.}~\bibnamefont {Chen}}, \bibinfo {author} {\bibfnamefont {H.-J.}\ \bibnamefont {Han}}, \bibinfo {author} {\bibfnamefont {C.}~\bibnamefont {Hinkle}}, \bibinfo {author} {\bibfnamefont {J.~A.}\ \bibnamefont {Robinson}}, \bibinfo {author} {\bibfnamefont {R.}~\bibnamefont {Sundararaman}},\ and\ \bibinfo {author} {\bibfnamefont {R.}~\bibnamefont {Torsi}},\ }\href {https://doi.org/10.1557/s43577-021-00192-3} {\bibfield  {journal} {\bibinfo  {journal} {MRS Bulletin}\ }\textbf {\bibinfo {volume} {46}},\ \bibinfo {pages} {959} (\bibinfo {year} {2021})}\BibitemShut {NoStop}%
\bibitem [{\citenamefont {van~der Veen}\ \emph {et~al.}(2016)\citenamefont {van~der Veen}, \citenamefont {Jourdan}, \citenamefont {Gonzalez}, \citenamefont {Wilson}, \citenamefont {Heylen}, \citenamefont {Pedreira}, \citenamefont {Struyf}, \citenamefont {Croes}, \citenamefont {Bömmels},\ and\ \citenamefont {Tőkei}}]{van_der_veen_barrierliner_2016}%
  \BibitemOpen
  \bibfield  {author} {\bibinfo {author} {\bibfnamefont {M.~H.}\ \bibnamefont {van~der Veen}}, \bibinfo {author} {\bibfnamefont {N.}~\bibnamefont {Jourdan}}, \bibinfo {author} {\bibfnamefont {V.~V.}\ \bibnamefont {Gonzalez}}, \bibinfo {author} {\bibfnamefont {C.~J.}\ \bibnamefont {Wilson}}, \bibinfo {author} {\bibfnamefont {N.}~\bibnamefont {Heylen}}, \bibinfo {author} {\bibfnamefont {O.~V.}\ \bibnamefont {Pedreira}}, \bibinfo {author} {\bibfnamefont {H.}~\bibnamefont {Struyf}}, \bibinfo {author} {\bibfnamefont {K.}~\bibnamefont {Croes}}, \bibinfo {author} {\bibfnamefont {J.}~\bibnamefont {Bömmels}},\ and\ \bibinfo {author} {\bibfnamefont {Z.}~\bibnamefont {Tőkei}},\ }in\ \href {https://doi.org/10.1109/IITC-AMC.2016.7507649} {\emph {\bibinfo {booktitle} {2016 {IEEE} {IITC}/{AMC}}}}\ (\bibinfo {year} {2016})\ pp.\ \bibinfo {pages} {28--30},\ \bibinfo {note} {iSSN: 2380-6338}\BibitemShut {NoStop}%
\bibitem [{\citenamefont {Chawla}\ and\ \citenamefont {Gall}(2009)}]{chawla_specular_2009}%
  \BibitemOpen
  \bibfield  {author} {\bibinfo {author} {\bibfnamefont {J.~S.}\ \bibnamefont {Chawla}}\ and\ \bibinfo {author} {\bibfnamefont {D.}~\bibnamefont {Gall}},\ }\href {https://doi.org/10.1063/1.3157271} {\bibfield  {journal} {\bibinfo  {journal} {Applied Physics Letters}\ }\textbf {\bibinfo {volume} {94}},\ \bibinfo {pages} {252101} (\bibinfo {year} {2009})}\BibitemShut {NoStop}%
\bibitem [{\citenamefont {Zheng}\ \emph {et~al.}(2014)\citenamefont {Zheng}, \citenamefont {Deng},\ and\ \citenamefont {Gall}}]{zheng_ni_2014}%
  \BibitemOpen
  \bibfield  {author} {\bibinfo {author} {\bibfnamefont {P.~Y.}\ \bibnamefont {Zheng}}, \bibinfo {author} {\bibfnamefont {R.~P.}\ \bibnamefont {Deng}},\ and\ \bibinfo {author} {\bibfnamefont {D.}~\bibnamefont {Gall}},\ }\href {https://doi.org/10.1063/1.4897009} {\bibfield  {journal} {\bibinfo  {journal} {Applied Physics Letters}\ }\textbf {\bibinfo {volume} {105}},\ \bibinfo {pages} {131603} (\bibinfo {year} {2014})}\BibitemShut {NoStop}%
\bibitem [{\citenamefont {Zheng}\ \emph {et~al.}(2016)\citenamefont {Zheng}, \citenamefont {Zhou},\ and\ \citenamefont {Gall}}]{zheng_electron_2016}%
  \BibitemOpen
  \bibfield  {author} {\bibinfo {author} {\bibfnamefont {P.}~\bibnamefont {Zheng}}, \bibinfo {author} {\bibfnamefont {T.}~\bibnamefont {Zhou}},\ and\ \bibinfo {author} {\bibfnamefont {D.}~\bibnamefont {Gall}},\ }\href {https://doi.org/10.1088/0268-1242/31/5/055005} {\bibfield  {journal} {\bibinfo  {journal} {Semiconductor Science and Technology}\ }\textbf {\bibinfo {volume} {31}},\ \bibinfo {pages} {055005} (\bibinfo {year} {2016})},\ \bibinfo {note} {publisher: IOP Publishing}\BibitemShut {NoStop}%
\bibitem [{\citenamefont {Barmak}\ \emph {et~al.}(2014)\citenamefont {Barmak}, \citenamefont {Darbal}, \citenamefont {Ganesh}, \citenamefont {Ferreira}, \citenamefont {Rickman}, \citenamefont {Sun}, \citenamefont {Yao}, \citenamefont {Warren},\ and\ \citenamefont {Coffey}}]{barmak_surface_2014}%
  \BibitemOpen
  \bibfield  {author} {\bibinfo {author} {\bibfnamefont {K.}~\bibnamefont {Barmak}}, \bibinfo {author} {\bibfnamefont {A.}~\bibnamefont {Darbal}}, \bibinfo {author} {\bibfnamefont {K.~J.}\ \bibnamefont {Ganesh}}, \bibinfo {author} {\bibfnamefont {P.~J.}\ \bibnamefont {Ferreira}}, \bibinfo {author} {\bibfnamefont {J.~M.}\ \bibnamefont {Rickman}}, \bibinfo {author} {\bibfnamefont {T.}~\bibnamefont {Sun}}, \bibinfo {author} {\bibfnamefont {B.}~\bibnamefont {Yao}}, \bibinfo {author} {\bibfnamefont {A.~P.}\ \bibnamefont {Warren}},\ and\ \bibinfo {author} {\bibfnamefont {K.~R.}\ \bibnamefont {Coffey}},\ }\href {https://doi.org/10.1116/1.4894453} {\bibfield  {journal} {\bibinfo  {journal} {Journal of Vacuum Science \& Technology A: Vacuum, Surfaces, and Films}\ }\textbf {\bibinfo {volume} {32}},\ \bibinfo {pages} {061503} (\bibinfo {year} {2014})}\BibitemShut {NoStop}%
\bibitem [{\citenamefont {César}\ \emph {et~al.}(2016)\citenamefont {César}, \citenamefont {Gall},\ and\ \citenamefont {Guo}}]{cesar_reducing_2016}%
  \BibitemOpen
  \bibfield  {author} {\bibinfo {author} {\bibfnamefont {M.}~\bibnamefont {César}}, \bibinfo {author} {\bibfnamefont {D.}~\bibnamefont {Gall}},\ and\ \bibinfo {author} {\bibfnamefont {H.}~\bibnamefont {Guo}},\ }\href {https://doi.org/10.1103/PhysRevApplied.5.054018} {\bibfield  {journal} {\bibinfo  {journal} {Physical Review Applied}\ }\textbf {\bibinfo {volume} {5}},\ \bibinfo {pages} {054018} (\bibinfo {year} {2016})}\BibitemShut {NoStop}%
\bibitem [{\citenamefont {César}\ \emph {et~al.}(2014)\citenamefont {César}, \citenamefont {Liu}, \citenamefont {Gall},\ and\ \citenamefont {Guo}}]{cesar_calculated_2014}%
  \BibitemOpen
  \bibfield  {author} {\bibinfo {author} {\bibfnamefont {M.}~\bibnamefont {César}}, \bibinfo {author} {\bibfnamefont {D.}~\bibnamefont {Liu}}, \bibinfo {author} {\bibfnamefont {D.}~\bibnamefont {Gall}},\ and\ \bibinfo {author} {\bibfnamefont {H.}~\bibnamefont {Guo}},\ }\href {https://doi.org/10.1103/PhysRevApplied.2.044007} {\bibfield  {journal} {\bibinfo  {journal} {Physical Review Applied}\ }\textbf {\bibinfo {volume} {2}},\ \bibinfo {pages} {044007} (\bibinfo {year} {2014})}\BibitemShut {NoStop}%
\bibitem [{\citenamefont {Kim}\ \emph {et~al.}(2010)\citenamefont {Kim}, \citenamefont {Zhang}, \citenamefont {Nicholson}, \citenamefont {Evans}, \citenamefont {Kulkarni}, \citenamefont {Radhakrishnan}, \citenamefont {Kenik},\ and\ \citenamefont {Li}}]{kim_large_2010}%
  \BibitemOpen
  \bibfield  {author} {\bibinfo {author} {\bibfnamefont {T.-H.}\ \bibnamefont {Kim}}, \bibinfo {author} {\bibfnamefont {X.-G.}\ \bibnamefont {Zhang}}, \bibinfo {author} {\bibfnamefont {D.~M.}\ \bibnamefont {Nicholson}}, \bibinfo {author} {\bibfnamefont {B.~M.}\ \bibnamefont {Evans}}, \bibinfo {author} {\bibfnamefont {N.~S.}\ \bibnamefont {Kulkarni}}, \bibinfo {author} {\bibfnamefont {B.}~\bibnamefont {Radhakrishnan}}, \bibinfo {author} {\bibfnamefont {E.~A.}\ \bibnamefont {Kenik}},\ and\ \bibinfo {author} {\bibfnamefont {A.-P.}\ \bibnamefont {Li}},\ }\href {https://doi.org/10.1021/nl101734h} {\bibfield  {journal} {\bibinfo  {journal} {Nano Letters}\ }\textbf {\bibinfo {volume} {10}},\ \bibinfo {pages} {3096} (\bibinfo {year} {2010})},\ \bibinfo {note} {publisher: American Chemical Society}\BibitemShut {NoStop}%
\bibitem [{\citenamefont {Rickman}\ and\ \citenamefont {Barmak}(2013)}]{rickman_simulation_2013}%
  \BibitemOpen
  \bibfield  {author} {\bibinfo {author} {\bibfnamefont {J.~M.}\ \bibnamefont {Rickman}}\ and\ \bibinfo {author} {\bibfnamefont {K.}~\bibnamefont {Barmak}},\ }\href {https://doi.org/10.1063/1.4823985} {\bibfield  {journal} {\bibinfo  {journal} {Journal of Applied Physics}\ }\textbf {\bibinfo {volume} {114}},\ \bibinfo {pages} {133703} (\bibinfo {year} {2013})}\BibitemShut {NoStop}%
\bibitem [{\citenamefont {Gall}(2020)}]{gall_search_2020}%
  \BibitemOpen
  \bibfield  {author} {\bibinfo {author} {\bibfnamefont {D.}~\bibnamefont {Gall}},\ }\href {https://doi.org/10.1063/1.5133671} {\bibfield  {journal} {\bibinfo  {journal} {Journal of Applied Physics}\ }\textbf {\bibinfo {volume} {127}},\ \bibinfo {pages} {050901} (\bibinfo {year} {2020})}\BibitemShut {NoStop}%
\bibitem [{\citenamefont {Dutta}\ \emph {et~al.}(2017{\natexlab{a}})\citenamefont {Dutta}, \citenamefont {Kundu}, \citenamefont {Wen}, \citenamefont {Jamieson}, \citenamefont {Croes}, \citenamefont {Gupta}, \citenamefont {Bömmels}, \citenamefont {Wilson}, \citenamefont {Adelmann},\ and\ \citenamefont {Tőkei}}]{dutta_ruthenium_2017}%
  \BibitemOpen
  \bibfield  {author} {\bibinfo {author} {\bibfnamefont {S.}~\bibnamefont {Dutta}}, \bibinfo {author} {\bibfnamefont {S.}~\bibnamefont {Kundu}}, \bibinfo {author} {\bibfnamefont {L.}~\bibnamefont {Wen}}, \bibinfo {author} {\bibfnamefont {G.}~\bibnamefont {Jamieson}}, \bibinfo {author} {\bibfnamefont {K.}~\bibnamefont {Croes}}, \bibinfo {author} {\bibfnamefont {A.}~\bibnamefont {Gupta}}, \bibinfo {author} {\bibfnamefont {J.}~\bibnamefont {Bömmels}}, \bibinfo {author} {\bibfnamefont {C.~J.}\ \bibnamefont {Wilson}}, \bibinfo {author} {\bibfnamefont {C.}~\bibnamefont {Adelmann}},\ and\ \bibinfo {author} {\bibfnamefont {Z.}~\bibnamefont {Tőkei}},\ }in\ \href {https://doi.org/10.1109/IITC-AMC.2017.7968937} {\emph {\bibinfo {booktitle} {2017 {IEEE} {IITC}}}}\ (\bibinfo {year} {2017})\ pp.\ \bibinfo {pages} {1--3},\ \bibinfo {note} {iSSN: 2380-6338}\BibitemShut {NoStop}%
\bibitem [{\citenamefont {Choi}\ \emph {et~al.}(2012)\citenamefont {Choi}, \citenamefont {Kim}, \citenamefont {Naveh}, \citenamefont {Chung}, \citenamefont {Warren}, \citenamefont {Nuhfer}, \citenamefont {Toney}, \citenamefont {Coffey},\ and\ \citenamefont {Barmak}}]{choi_electron_2012}%
  \BibitemOpen
  \bibfield  {author} {\bibinfo {author} {\bibfnamefont {D.}~\bibnamefont {Choi}}, \bibinfo {author} {\bibfnamefont {C.~S.}\ \bibnamefont {Kim}}, \bibinfo {author} {\bibfnamefont {D.}~\bibnamefont {Naveh}}, \bibinfo {author} {\bibfnamefont {S.}~\bibnamefont {Chung}}, \bibinfo {author} {\bibfnamefont {A.~P.}\ \bibnamefont {Warren}}, \bibinfo {author} {\bibfnamefont {N.~T.}\ \bibnamefont {Nuhfer}}, \bibinfo {author} {\bibfnamefont {M.~F.}\ \bibnamefont {Toney}}, \bibinfo {author} {\bibfnamefont {K.~R.}\ \bibnamefont {Coffey}},\ and\ \bibinfo {author} {\bibfnamefont {K.}~\bibnamefont {Barmak}},\ }\href {https://doi.org/10.1103/PhysRevB.86.045432} {\bibfield  {journal} {\bibinfo  {journal} {Physical Review B}\ }\textbf {\bibinfo {volume} {86}},\ \bibinfo {pages} {045432} (\bibinfo {year} {2012})}\BibitemShut {NoStop}%
\bibitem [{\citenamefont {Kamineni}\ \emph {et~al.}(2016)\citenamefont {Kamineni}, \citenamefont {Raymond}, \citenamefont {Siddiqui}, \citenamefont {Mont}, \citenamefont {Tsai}, \citenamefont {Niu}, \citenamefont {Labonte}, \citenamefont {Labelle}, \citenamefont {Fan}, \citenamefont {Peethala}, \citenamefont {Adusumilli}, \citenamefont {Patlolla}, \citenamefont {Priyadarshini}, \citenamefont {Mignot}, \citenamefont {Carr}, \citenamefont {Pancharatnam}, \citenamefont {Shearer}, \citenamefont {Surisetty}, \citenamefont {Arnold}, \citenamefont {Canaperi}, \citenamefont {Haran}, \citenamefont {Jagannathan}, \citenamefont {Chafik},\ and\ \citenamefont {L'Herron}}]{kamineni_tungsten_2016}%
  \BibitemOpen
  \bibfield  {author} {\bibinfo {author} {\bibfnamefont {V.}~\bibnamefont {Kamineni}}, \bibinfo {author} {\bibfnamefont {M.}~\bibnamefont {Raymond}}, \bibinfo {author} {\bibfnamefont {S.}~\bibnamefont {Siddiqui}}, \bibinfo {author} {\bibfnamefont {F.}~\bibnamefont {Mont}}, \bibinfo {author} {\bibfnamefont {S.}~\bibnamefont {Tsai}}, \bibinfo {author} {\bibfnamefont {C.}~\bibnamefont {Niu}}, \bibinfo {author} {\bibfnamefont {A.}~\bibnamefont {Labonte}}, \bibinfo {author} {\bibfnamefont {C.}~\bibnamefont {Labelle}}, \bibinfo {author} {\bibfnamefont {S.}~\bibnamefont {Fan}}, \bibinfo {author} {\bibfnamefont {B.}~\bibnamefont {Peethala}}, \bibinfo {author} {\bibfnamefont {P.}~\bibnamefont {Adusumilli}}, \bibinfo {author} {\bibfnamefont {R.}~\bibnamefont {Patlolla}}, \bibinfo {author} {\bibfnamefont {D.}~\bibnamefont {Priyadarshini}}, \bibinfo {author} {\bibfnamefont {Y.}~\bibnamefont {Mignot}}, \bibinfo {author} {\bibfnamefont {A.}~\bibnamefont {Carr}}, \bibinfo {author} {\bibfnamefont {S.}~\bibnamefont
  {Pancharatnam}}, \bibinfo {author} {\bibfnamefont {J.}~\bibnamefont {Shearer}}, \bibinfo {author} {\bibfnamefont {C.}~\bibnamefont {Surisetty}}, \bibinfo {author} {\bibfnamefont {J.}~\bibnamefont {Arnold}}, \bibinfo {author} {\bibfnamefont {D.}~\bibnamefont {Canaperi}}, \bibinfo {author} {\bibfnamefont {B.}~\bibnamefont {Haran}}, \bibinfo {author} {\bibfnamefont {H.}~\bibnamefont {Jagannathan}}, \bibinfo {author} {\bibfnamefont {F.}~\bibnamefont {Chafik}},\ and\ \bibinfo {author} {\bibfnamefont {B.}~\bibnamefont {L'Herron}},\ }in\ \href {https://doi.org/10.1109/IITC-AMC.2016.7507698} {\emph {\bibinfo {booktitle} {2016 {IEEE} {IITC}/{AMC}}}}\ (\bibinfo {year} {2016})\ pp.\ \bibinfo {pages} {105--107},\ \bibinfo {note} {iSSN: 2380-6338}\BibitemShut {NoStop}%
\bibitem [{\citenamefont {Chawla}\ \emph {et~al.}(2016)\citenamefont {Chawla}, \citenamefont {Sung}, \citenamefont {Bojarski}, \citenamefont {Carver}, \citenamefont {Chandhok}, \citenamefont {Chebiam}, \citenamefont {Clarke}, \citenamefont {Harmes}, \citenamefont {Jezewski}, \citenamefont {Kobrinski}, \citenamefont {Krist}, \citenamefont {Mayeh}, \citenamefont {Turkot},\ and\ \citenamefont {Yoo}}]{chawla_resistance_2016}%
  \BibitemOpen
  \bibfield  {author} {\bibinfo {author} {\bibfnamefont {J.~S.}\ \bibnamefont {Chawla}}, \bibinfo {author} {\bibfnamefont {S.~H.}\ \bibnamefont {Sung}}, \bibinfo {author} {\bibfnamefont {S.~A.}\ \bibnamefont {Bojarski}}, \bibinfo {author} {\bibfnamefont {C.~T.}\ \bibnamefont {Carver}}, \bibinfo {author} {\bibfnamefont {M.}~\bibnamefont {Chandhok}}, \bibinfo {author} {\bibfnamefont {R.~V.}\ \bibnamefont {Chebiam}}, \bibinfo {author} {\bibfnamefont {J.~S.}\ \bibnamefont {Clarke}}, \bibinfo {author} {\bibfnamefont {M.}~\bibnamefont {Harmes}}, \bibinfo {author} {\bibfnamefont {C.~J.}\ \bibnamefont {Jezewski}}, \bibinfo {author} {\bibfnamefont {M.~J.}\ \bibnamefont {Kobrinski}}, \bibinfo {author} {\bibfnamefont {B.~J.}\ \bibnamefont {Krist}}, \bibinfo {author} {\bibfnamefont {M.}~\bibnamefont {Mayeh}}, \bibinfo {author} {\bibfnamefont {R.}~\bibnamefont {Turkot}},\ and\ \bibinfo {author} {\bibfnamefont {H.~J.}\ \bibnamefont {Yoo}},\ }in\ \href {https://doi.org/10.1109/IITC-AMC.2016.7507682} {\emph {\bibinfo
  {booktitle} {2016 {IEEE} {IITC}/{AMC}}}}\ (\bibinfo {year} {2016})\ pp.\ \bibinfo {pages} {63--65},\ \bibinfo {note} {iSSN: 2380-6338}\BibitemShut {NoStop}%
\bibitem [{\citenamefont {Zhang}\ \emph {et~al.}(2016)\citenamefont {Zhang}, \citenamefont {Huang}, \citenamefont {Patlolla}, \citenamefont {Wang}, \citenamefont {Mont}, \citenamefont {Li}, \citenamefont {Hu}, \citenamefont {Liniger}, \citenamefont {McLaughlin}, \citenamefont {Labelle}, \citenamefont {Ryan}, \citenamefont {Canaperi}, \citenamefont {Spooner}, \citenamefont {Bonilla},\ and\ \citenamefont {Edelstein}}]{zhang_ruthenium_2016}%
  \BibitemOpen
  \bibfield  {author} {\bibinfo {author} {\bibfnamefont {X.}~\bibnamefont {Zhang}}, \bibinfo {author} {\bibfnamefont {H.}~\bibnamefont {Huang}}, \bibinfo {author} {\bibfnamefont {R.}~\bibnamefont {Patlolla}}, \bibinfo {author} {\bibfnamefont {W.}~\bibnamefont {Wang}}, \bibinfo {author} {\bibfnamefont {F.~W.}\ \bibnamefont {Mont}}, \bibinfo {author} {\bibfnamefont {J.}~\bibnamefont {Li}}, \bibinfo {author} {\bibfnamefont {C.-K.}\ \bibnamefont {Hu}}, \bibinfo {author} {\bibfnamefont {E.~G.}\ \bibnamefont {Liniger}}, \bibinfo {author} {\bibfnamefont {P.~S.}\ \bibnamefont {McLaughlin}}, \bibinfo {author} {\bibfnamefont {C.}~\bibnamefont {Labelle}}, \bibinfo {author} {\bibfnamefont {E.~T.}\ \bibnamefont {Ryan}}, \bibinfo {author} {\bibfnamefont {D.}~\bibnamefont {Canaperi}}, \bibinfo {author} {\bibfnamefont {T.}~\bibnamefont {Spooner}}, \bibinfo {author} {\bibfnamefont {G.}~\bibnamefont {Bonilla}},\ and\ \bibinfo {author} {\bibfnamefont {D.}~\bibnamefont {Edelstein}},\ }in\ \href
  {https://doi.org/10.1109/IITC-AMC.2016.7507650} {\emph {\bibinfo {booktitle} {2016 {IEEE} {IITC}/{AMC}}}}\ (\bibinfo {year} {2016})\ pp.\ \bibinfo {pages} {31--33},\ \bibinfo {note} {iSSN: 2380-6338}\BibitemShut {NoStop}%
\bibitem [{\citenamefont {Kelly}\ \emph {et~al.}(2016)\citenamefont {Kelly}, \citenamefont {Chen}, \citenamefont {Huang}, \citenamefont {Hu}, \citenamefont {Liniger}, \citenamefont {Patlolla}, \citenamefont {Peethala}, \citenamefont {Adusumilli}, \citenamefont {Shobha}, \citenamefont {Nogami}, \citenamefont {Spooner}, \citenamefont {Huang}, \citenamefont {Edelstein}, \citenamefont {Canaperi}, \citenamefont {Kamineni}, \citenamefont {Mont},\ and\ \citenamefont {Siddiqui}}]{kelly_experimental_2016}%
  \BibitemOpen
  \bibfield  {author} {\bibinfo {author} {\bibfnamefont {J.}~\bibnamefont {Kelly}}, \bibinfo {author} {\bibfnamefont {J.~H.-C.}\ \bibnamefont {Chen}}, \bibinfo {author} {\bibfnamefont {H.}~\bibnamefont {Huang}}, \bibinfo {author} {\bibfnamefont {C.~K.}\ \bibnamefont {Hu}}, \bibinfo {author} {\bibfnamefont {E.}~\bibnamefont {Liniger}}, \bibinfo {author} {\bibfnamefont {R.}~\bibnamefont {Patlolla}}, \bibinfo {author} {\bibfnamefont {B.}~\bibnamefont {Peethala}}, \bibinfo {author} {\bibfnamefont {P.}~\bibnamefont {Adusumilli}}, \bibinfo {author} {\bibfnamefont {H.}~\bibnamefont {Shobha}}, \bibinfo {author} {\bibfnamefont {T.}~\bibnamefont {Nogami}}, \bibinfo {author} {\bibfnamefont {T.}~\bibnamefont {Spooner}}, \bibinfo {author} {\bibfnamefont {E.}~\bibnamefont {Huang}}, \bibinfo {author} {\bibfnamefont {D.}~\bibnamefont {Edelstein}}, \bibinfo {author} {\bibfnamefont {D.}~\bibnamefont {Canaperi}}, \bibinfo {author} {\bibfnamefont {V.}~\bibnamefont {Kamineni}}, \bibinfo {author} {\bibfnamefont {F.}~\bibnamefont
  {Mont}},\ and\ \bibinfo {author} {\bibfnamefont {S.}~\bibnamefont {Siddiqui}},\ }in\ \href {https://doi.org/10.1109/IITC-AMC.2016.7507673} {\emph {\bibinfo {booktitle} {2016 {IEEE} {IITC}/{AMC}}}}\ (\bibinfo {year} {2016})\ pp.\ \bibinfo {pages} {40--42},\ \bibinfo {note} {iSSN: 2380-6338}\BibitemShut {NoStop}%
\bibitem [{\citenamefont {Wen}\ \emph {et~al.}(2016)\citenamefont {Wen}, \citenamefont {Adelmann}, \citenamefont {Pedreira}, \citenamefont {Dutta}, \citenamefont {Popovici}, \citenamefont {Briggs}, \citenamefont {Heylen}, \citenamefont {Vanstreels}, \citenamefont {Wilson}, \citenamefont {Van~Elshocht}, \citenamefont {Croes}, \citenamefont {Bömmels},\ and\ \citenamefont {Tőkei}}]{wen_ruthenium_2016}%
  \BibitemOpen
  \bibfield  {author} {\bibinfo {author} {\bibfnamefont {L.~G.}\ \bibnamefont {Wen}}, \bibinfo {author} {\bibfnamefont {C.}~\bibnamefont {Adelmann}}, \bibinfo {author} {\bibfnamefont {O.~V.}\ \bibnamefont {Pedreira}}, \bibinfo {author} {\bibfnamefont {S.}~\bibnamefont {Dutta}}, \bibinfo {author} {\bibfnamefont {M.}~\bibnamefont {Popovici}}, \bibinfo {author} {\bibfnamefont {B.}~\bibnamefont {Briggs}}, \bibinfo {author} {\bibfnamefont {N.}~\bibnamefont {Heylen}}, \bibinfo {author} {\bibfnamefont {K.}~\bibnamefont {Vanstreels}}, \bibinfo {author} {\bibfnamefont {C.~J.}\ \bibnamefont {Wilson}}, \bibinfo {author} {\bibfnamefont {S.}~\bibnamefont {Van~Elshocht}}, \bibinfo {author} {\bibfnamefont {K.}~\bibnamefont {Croes}}, \bibinfo {author} {\bibfnamefont {J.}~\bibnamefont {Bömmels}},\ and\ \bibinfo {author} {\bibfnamefont {Z.}~\bibnamefont {Tőkei}},\ }in\ \href {https://doi.org/10.1109/IITC-AMC.2016.7507651} {\emph {\bibinfo {booktitle} {2016 {IEEE} {IITC}/{AMC}}}}\ (\bibinfo {year} {2016})\ pp.\ \bibinfo
  {pages} {34--36},\ \bibinfo {note} {iSSN: 2380-6338}\BibitemShut {NoStop}%
\bibitem [{\citenamefont {Dutta}\ \emph {et~al.}(2017{\natexlab{b}})\citenamefont {Dutta}, \citenamefont {Sankaran}, \citenamefont {Moors}, \citenamefont {Pourtois}, \citenamefont {Van~Elshocht}, \citenamefont {Bömmels}, \citenamefont {Vandervorst}, \citenamefont {Tőkei},\ and\ \citenamefont {Adelmann}}]{dutta_thickness_2017}%
  \BibitemOpen
  \bibfield  {author} {\bibinfo {author} {\bibfnamefont {S.}~\bibnamefont {Dutta}}, \bibinfo {author} {\bibfnamefont {K.}~\bibnamefont {Sankaran}}, \bibinfo {author} {\bibfnamefont {K.}~\bibnamefont {Moors}}, \bibinfo {author} {\bibfnamefont {G.}~\bibnamefont {Pourtois}}, \bibinfo {author} {\bibfnamefont {S.}~\bibnamefont {Van~Elshocht}}, \bibinfo {author} {\bibfnamefont {J.}~\bibnamefont {Bömmels}}, \bibinfo {author} {\bibfnamefont {W.}~\bibnamefont {Vandervorst}}, \bibinfo {author} {\bibfnamefont {Z.}~\bibnamefont {Tőkei}},\ and\ \bibinfo {author} {\bibfnamefont {C.}~\bibnamefont {Adelmann}},\ }\href {https://doi.org/10.1063/1.4992089} {\bibfield  {journal} {\bibinfo  {journal} {Journal of Applied Physics}\ }\textbf {\bibinfo {volume} {122}},\ \bibinfo {pages} {025107} (\bibinfo {year} {2017}{\natexlab{b}})}\BibitemShut {NoStop}%
\bibitem [{\citenamefont {Zhang}\ \emph {et~al.}(2017)\citenamefont {Zhang}, \citenamefont {Huang}, \citenamefont {Patlolla}, \citenamefont {Mont}, \citenamefont {Lin}, \citenamefont {Raymond}, \citenamefont {Labelle}, \citenamefont {Ryan}, \citenamefont {Canaperi}, \citenamefont {Standaert}, \citenamefont {Spooner}, \citenamefont {Bonilla},\ and\ \citenamefont {Edelstein}}]{zhang_methods_2017}%
  \BibitemOpen
  \bibfield  {author} {\bibinfo {author} {\bibfnamefont {X.}~\bibnamefont {Zhang}}, \bibinfo {author} {\bibfnamefont {H.}~\bibnamefont {Huang}}, \bibinfo {author} {\bibfnamefont {R.}~\bibnamefont {Patlolla}}, \bibinfo {author} {\bibfnamefont {F.~W.}\ \bibnamefont {Mont}}, \bibinfo {author} {\bibfnamefont {X.}~\bibnamefont {Lin}}, \bibinfo {author} {\bibfnamefont {M.}~\bibnamefont {Raymond}}, \bibinfo {author} {\bibfnamefont {C.}~\bibnamefont {Labelle}}, \bibinfo {author} {\bibfnamefont {E.~T.}\ \bibnamefont {Ryan}}, \bibinfo {author} {\bibfnamefont {D.}~\bibnamefont {Canaperi}}, \bibinfo {author} {\bibfnamefont {T.~E.}\ \bibnamefont {Standaert}}, \bibinfo {author} {\bibfnamefont {T.}~\bibnamefont {Spooner}}, \bibinfo {author} {\bibfnamefont {G.}~\bibnamefont {Bonilla}},\ and\ \bibinfo {author} {\bibfnamefont {D.}~\bibnamefont {Edelstein}},\ }in\ \href {https://doi.org/10.1109/IITC-AMC.2017.7968941} {\emph {\bibinfo {booktitle} {2017 {IEEE} {IITC}}}}\ (\bibinfo {year} {2017})\ pp.\ \bibinfo {pages} {1--3},\
  \bibinfo {note} {iSSN: 2380-6338}\BibitemShut {NoStop}%
\bibitem [{\citenamefont {Hu}\ \emph {et~al.}(2017)\citenamefont {Hu}, \citenamefont {Kelly}, \citenamefont {Chen}, \citenamefont {Huang}, \citenamefont {Ostrovski}, \citenamefont {Patlolla}, \citenamefont {Peethala}, \citenamefont {Adusumilli}, \citenamefont {Spooner}, \citenamefont {Gignac}, \citenamefont {Bruley}, \citenamefont {Breslin}, \citenamefont {Cohen}, \citenamefont {Lian}, \citenamefont {Ali}, \citenamefont {Long}, \citenamefont {Hornicek}, \citenamefont {Kane}, \citenamefont {Kamineni}, \citenamefont {Zhang}, \citenamefont {Mont},\ and\ \citenamefont {Siddiqui}}]{hu_electromigration_2017}%
  \BibitemOpen
  \bibfield  {author} {\bibinfo {author} {\bibfnamefont {C.-K.}\ \bibnamefont {Hu}}, \bibinfo {author} {\bibfnamefont {J.}~\bibnamefont {Kelly}}, \bibinfo {author} {\bibfnamefont {J.~H.-C.}\ \bibnamefont {Chen}}, \bibinfo {author} {\bibfnamefont {H.}~\bibnamefont {Huang}}, \bibinfo {author} {\bibfnamefont {Y.}~\bibnamefont {Ostrovski}}, \bibinfo {author} {\bibfnamefont {R.}~\bibnamefont {Patlolla}}, \bibinfo {author} {\bibfnamefont {B.}~\bibnamefont {Peethala}}, \bibinfo {author} {\bibfnamefont {P.}~\bibnamefont {Adusumilli}}, \bibinfo {author} {\bibfnamefont {T.}~\bibnamefont {Spooner}}, \bibinfo {author} {\bibfnamefont {L.~M.}\ \bibnamefont {Gignac}}, \bibinfo {author} {\bibfnamefont {J.}~\bibnamefont {Bruley}}, \bibinfo {author} {\bibfnamefont {C.}~\bibnamefont {Breslin}}, \bibinfo {author} {\bibfnamefont {S.~A.}\ \bibnamefont {Cohen}}, \bibinfo {author} {\bibfnamefont {G.}~\bibnamefont {Lian}}, \bibinfo {author} {\bibfnamefont {M.}~\bibnamefont {Ali}}, \bibinfo {author} {\bibfnamefont {R.}~\bibnamefont
  {Long}}, \bibinfo {author} {\bibfnamefont {G.}~\bibnamefont {Hornicek}}, \bibinfo {author} {\bibfnamefont {T.}~\bibnamefont {Kane}}, \bibinfo {author} {\bibfnamefont {V.}~\bibnamefont {Kamineni}}, \bibinfo {author} {\bibfnamefont {X.}~\bibnamefont {Zhang}}, \bibinfo {author} {\bibfnamefont {F.}~\bibnamefont {Mont}},\ and\ \bibinfo {author} {\bibfnamefont {S.}~\bibnamefont {Siddiqui}},\ }in\ \href {https://doi.org/10.1109/IITC-AMC.2017.7968977} {\emph {\bibinfo {booktitle} {2017 {IEEE} {IITC}}}}\ (\bibinfo {year} {2017})\ pp.\ \bibinfo {pages} {1--3},\ \bibinfo {note} {iSSN: 2380-6338}\BibitemShut {NoStop}%
\bibitem [{\citenamefont {Gall}(2016)}]{gall_electron_2016}%
  \BibitemOpen
  \bibfield  {author} {\bibinfo {author} {\bibfnamefont {D.}~\bibnamefont {Gall}},\ }\href {https://doi.org/10.1063/1.4942216} {\bibfield  {journal} {\bibinfo  {journal} {Journal of Applied Physics}\ }\textbf {\bibinfo {volume} {119}},\ \bibinfo {pages} {085101} (\bibinfo {year} {2016})}\BibitemShut {NoStop}%
\bibitem [{\citenamefont {Gall}(2018)}]{gall_metals_2018}%
  \BibitemOpen
  \bibfield  {author} {\bibinfo {author} {\bibfnamefont {D.}~\bibnamefont {Gall}},\ }in\ \href {https://doi.org/10.1109/IITC.2018.8456810} {\emph {\bibinfo {booktitle} {2018 {IEEE} {IITC}}}}\ (\bibinfo {year} {2018})\ pp.\ \bibinfo {pages} {157--159},\ \bibinfo {note} {iSSN: 2380-6338}\BibitemShut {NoStop}%
\bibitem [{\citenamefont {Han}\ \emph {et~al.}(2021)\citenamefont {Han}, \citenamefont {Liu},\ and\ \citenamefont {Cha}}]{han_1d_2021}%
  \BibitemOpen
  \bibfield  {author} {\bibinfo {author} {\bibfnamefont {H.~J.}\ \bibnamefont {Han}}, \bibinfo {author} {\bibfnamefont {P.}~\bibnamefont {Liu}},\ and\ \bibinfo {author} {\bibfnamefont {J.~J.}\ \bibnamefont {Cha}},\ }\href {https://doi.org/10.1016/j.matt.2021.05.020} {\bibfield  {journal} {\bibinfo  {journal} {Matter}\ }\textbf {\bibinfo {volume} {4}},\ \bibinfo {pages} {2596} (\bibinfo {year} {2021})}\BibitemShut {NoStop}%
\bibitem [{\citenamefont {Zhang}\ \emph {et~al.}(2019)\citenamefont {Zhang}, \citenamefont {Ni}, \citenamefont {Zhang}, \citenamefont {Yuan}, \citenamefont {Liu}, \citenamefont {Zou}, \citenamefont {Liao}, \citenamefont {Du}, \citenamefont {Narayan}, \citenamefont {Zhang}, \citenamefont {Gu}, \citenamefont {Zhu}, \citenamefont {Pi}, \citenamefont {Sanvito}, \citenamefont {Han}, \citenamefont {Zou}, \citenamefont {Shi}, \citenamefont {Wan}, \citenamefont {Savrasov},\ and\ \citenamefont {Xiu}}]{zhang_ultrahigh_2019}%
  \BibitemOpen
  \bibfield  {author} {\bibinfo {author} {\bibfnamefont {C.}~\bibnamefont {Zhang}}, \bibinfo {author} {\bibfnamefont {Z.}~\bibnamefont {Ni}}, \bibinfo {author} {\bibfnamefont {J.}~\bibnamefont {Zhang}}, \bibinfo {author} {\bibfnamefont {X.}~\bibnamefont {Yuan}}, \bibinfo {author} {\bibfnamefont {Y.}~\bibnamefont {Liu}}, \bibinfo {author} {\bibfnamefont {Y.}~\bibnamefont {Zou}}, \bibinfo {author} {\bibfnamefont {Z.}~\bibnamefont {Liao}}, \bibinfo {author} {\bibfnamefont {Y.}~\bibnamefont {Du}}, \bibinfo {author} {\bibfnamefont {A.}~\bibnamefont {Narayan}}, \bibinfo {author} {\bibfnamefont {H.}~\bibnamefont {Zhang}}, \bibinfo {author} {\bibfnamefont {T.}~\bibnamefont {Gu}}, \bibinfo {author} {\bibfnamefont {X.}~\bibnamefont {Zhu}}, \bibinfo {author} {\bibfnamefont {L.}~\bibnamefont {Pi}}, \bibinfo {author} {\bibfnamefont {S.}~\bibnamefont {Sanvito}}, \bibinfo {author} {\bibfnamefont {X.}~\bibnamefont {Han}}, \bibinfo {author} {\bibfnamefont {J.}~\bibnamefont {Zou}}, \bibinfo {author} {\bibfnamefont
  {Y.}~\bibnamefont {Shi}}, \bibinfo {author} {\bibfnamefont {X.}~\bibnamefont {Wan}}, \bibinfo {author} {\bibfnamefont {S.~Y.}\ \bibnamefont {Savrasov}},\ and\ \bibinfo {author} {\bibfnamefont {F.}~\bibnamefont {Xiu}},\ }\href {https://doi.org/10.1038/s41563-019-0320-9} {\bibfield  {journal} {\bibinfo  {journal} {Nature Materials}\ }\textbf {\bibinfo {volume} {18}},\ \bibinfo {pages} {482} (\bibinfo {year} {2019})},\ \bibinfo {note} {publisher: Nature Publishing Group}\BibitemShut {NoStop}%
\bibitem [{\citenamefont {Lien}\ \emph {et~al.}(2023)\citenamefont {Lien}, \citenamefont {Garate}, \citenamefont {Bajpai}, \citenamefont {Huang}, \citenamefont {Hsu}, \citenamefont {Tu}, \citenamefont {Lanzillo}, \citenamefont {Bansil}, \citenamefont {Chang}, \citenamefont {Liang}, \citenamefont {Lin},\ and\ \citenamefont {Chen}}]{lien_unconventional_2023}%
  \BibitemOpen
  \bibfield  {author} {\bibinfo {author} {\bibfnamefont {S.-W.}\ \bibnamefont {Lien}}, \bibinfo {author} {\bibfnamefont {I.}~\bibnamefont {Garate}}, \bibinfo {author} {\bibfnamefont {U.}~\bibnamefont {Bajpai}}, \bibinfo {author} {\bibfnamefont {C.-Y.}\ \bibnamefont {Huang}}, \bibinfo {author} {\bibfnamefont {C.-H.}\ \bibnamefont {Hsu}}, \bibinfo {author} {\bibfnamefont {Y.-H.}\ \bibnamefont {Tu}}, \bibinfo {author} {\bibfnamefont {N.~A.}\ \bibnamefont {Lanzillo}}, \bibinfo {author} {\bibfnamefont {A.}~\bibnamefont {Bansil}}, \bibinfo {author} {\bibfnamefont {T.-R.}\ \bibnamefont {Chang}}, \bibinfo {author} {\bibfnamefont {G.}~\bibnamefont {Liang}}, \bibinfo {author} {\bibfnamefont {H.}~\bibnamefont {Lin}},\ and\ \bibinfo {author} {\bibfnamefont {C.-T.}\ \bibnamefont {Chen}},\ }\href {https://doi.org/10.1038/s41535-022-00535-6} {\bibfield  {journal} {\bibinfo  {journal} {npj Quantum Materials}\ }\textbf {\bibinfo {volume} {8}},\ \bibinfo {pages} {1} (\bibinfo {year} {2023})},\ \bibinfo {note} {publisher:
  Nature Publishing Group}\BibitemShut {NoStop}%
\bibitem [{\citenamefont {Rocchino}\ \emph {et~al.}(2024)\citenamefont {Rocchino}, \citenamefont {Molinari}, \citenamefont {Kladaric}, \citenamefont {Balduini}, \citenamefont {Schmid}, \citenamefont {Sousa}, \citenamefont {Bruley}, \citenamefont {Bui}, \citenamefont {Gotsmann},\ and\ \citenamefont {Zota}}]{rocchino_unconventional_2024}%
  \BibitemOpen
  \bibfield  {author} {\bibinfo {author} {\bibfnamefont {L.}~\bibnamefont {Rocchino}}, \bibinfo {author} {\bibfnamefont {A.}~\bibnamefont {Molinari}}, \bibinfo {author} {\bibfnamefont {I.}~\bibnamefont {Kladaric}}, \bibinfo {author} {\bibfnamefont {F.}~\bibnamefont {Balduini}}, \bibinfo {author} {\bibfnamefont {H.}~\bibnamefont {Schmid}}, \bibinfo {author} {\bibfnamefont {M.}~\bibnamefont {Sousa}}, \bibinfo {author} {\bibfnamefont {J.}~\bibnamefont {Bruley}}, \bibinfo {author} {\bibfnamefont {H.}~\bibnamefont {Bui}}, \bibinfo {author} {\bibfnamefont {B.}~\bibnamefont {Gotsmann}},\ and\ \bibinfo {author} {\bibfnamefont {C.~B.}\ \bibnamefont {Zota}},\ }\href {https://doi.org/10.1038/s41598-024-71614-w} {\bibfield  {journal} {\bibinfo  {journal} {Scientific Reports}\ }\textbf {\bibinfo {volume} {14}},\ \bibinfo {pages} {20608} (\bibinfo {year} {2024})},\ \bibinfo {note} {publisher: Nature Publishing Group}\BibitemShut {NoStop}%
\bibitem [{\citenamefont {Kumar}\ \emph {et~al.}(2022)\citenamefont {Kumar}, \citenamefont {Multunas}, \citenamefont {Defay}, \citenamefont {Gall},\ and\ \citenamefont {Sundararaman}}]{kumar_ultralow_2022}%
  \BibitemOpen
  \bibfield  {author} {\bibinfo {author} {\bibfnamefont {S.}~\bibnamefont {Kumar}}, \bibinfo {author} {\bibfnamefont {C.}~\bibnamefont {Multunas}}, \bibinfo {author} {\bibfnamefont {B.}~\bibnamefont {Defay}}, \bibinfo {author} {\bibfnamefont {D.}~\bibnamefont {Gall}},\ and\ \bibinfo {author} {\bibfnamefont {R.}~\bibnamefont {Sundararaman}},\ }\href {https://doi.org/10.1103/PhysRevMaterials.6.085002} {\bibfield  {journal} {\bibinfo  {journal} {Physical Review Materials}\ }\textbf {\bibinfo {volume} {6}},\ \bibinfo {pages} {085002} (\bibinfo {year} {2022})}\BibitemShut {NoStop}%
\bibitem [{\citenamefont {Zheng}\ and\ \citenamefont {Gall}(2017)}]{zheng_anisotropic_2017}%
  \BibitemOpen
  \bibfield  {author} {\bibinfo {author} {\bibfnamefont {P.}~\bibnamefont {Zheng}}\ and\ \bibinfo {author} {\bibfnamefont {D.}~\bibnamefont {Gall}},\ }\href {https://doi.org/10.1063/1.5004118} {\bibfield  {journal} {\bibinfo  {journal} {Journal of Applied Physics}\ }\textbf {\bibinfo {volume} {122}},\ \bibinfo {pages} {135301} (\bibinfo {year} {2017})}\BibitemShut {NoStop}%
\bibitem [{\citenamefont {Hashimoto}\ and\ \citenamefont {Ueda}(1998)}]{hashimoto_anisotropy_1998}%
  \BibitemOpen
  \bibfield  {author} {\bibinfo {author} {\bibfnamefont {E.}~\bibnamefont {Hashimoto}}\ and\ \bibinfo {author} {\bibfnamefont {Y.}~\bibnamefont {Ueda}},\ }\href {https://doi.org/10.1088/0953-8984/10/30/012} {\bibfield  {journal} {\bibinfo  {journal} {Journal of Physics: Condensed Matter}\ }\textbf {\bibinfo {volume} {10}},\ \bibinfo {pages} {6727} (\bibinfo {year} {1998})}\BibitemShut {NoStop}%
\bibitem [{\citenamefont {Schoenberg}(2009)}]{schoenberg_magnetic_2009}%
  \BibitemOpen
  \bibfield  {author} {\bibinfo {author} {\bibfnamefont {D.}~\bibnamefont {Schoenberg}},\ }\bibfield  {journal} {\bibinfo  {journal} {Magnetic Oscillations in Metals, by D. Shoenberg, Cambridge, UK: Cambridge University Press, 2009}\ }\href {https://doi.org/10.1017/CBO9780511897870} {10.1017/CBO9780511897870} (\bibinfo {year} {2009})\BibitemShut {NoStop}%
\bibitem [{\citenamefont {Onsager}(1952)}]{onsager_interpretation_1952}%
  \BibitemOpen
  \bibfield  {author} {\bibinfo {author} {\bibfnamefont {L.}~\bibnamefont {Onsager}},\ }\bibfield  {journal} {\bibinfo  {journal} {The London, Edinburgh, and Dublin Philosophical Magazine and Journal of Science}\ }\href {https://doi.org/10.1080/14786440908521019} {10.1080/14786440908521019} (\bibinfo {year} {1952}),\ \bibinfo {note} {publisher: Taylor \& Francis Group}\BibitemShut {NoStop}%
\bibitem [{\citenamefont {Roth}(1966)}]{roth_semiclassical_1966}%
  \BibitemOpen
  \bibfield  {author} {\bibinfo {author} {\bibfnamefont {L.~M.}\ \bibnamefont {Roth}},\ }\href {https://doi.org/10.1103/PhysRev.145.434} {\bibfield  {journal} {\bibinfo  {journal} {Physical Review}\ }\textbf {\bibinfo {volume} {145}},\ \bibinfo {pages} {434} (\bibinfo {year} {1966})},\ \bibinfo {note} {number: 2 Publisher: American Physical Society}\BibitemShut {NoStop}%
\bibitem [{\citenamefont {Ashcroft}(1976)}]{ashcroft_solid_1976}%
  \BibitemOpen
  \bibfield  {author} {\bibinfo {author} {\bibfnamefont {N.~W.}\ \bibnamefont {Ashcroft}},\ }\href@noop {} {\emph {\bibinfo {title} {Solid state physics}}}\ (\bibinfo  {publisher} {Thomson},\ \bibinfo {address} {Singapore},\ \bibinfo {year} {1976})\BibitemShut {NoStop}%
\bibitem [{\citenamefont {Schindler}\ \emph {et~al.}(2020)\citenamefont {Schindler}, \citenamefont {Noky}, \citenamefont {Schmidt}, \citenamefont {Felser}, \citenamefont {Wosnitza},\ and\ \citenamefont {Gooth}}]{schindler_effect_2020}%
  \BibitemOpen
  \bibfield  {author} {\bibinfo {author} {\bibfnamefont {C.}~\bibnamefont {Schindler}}, \bibinfo {author} {\bibfnamefont {J.}~\bibnamefont {Noky}}, \bibinfo {author} {\bibfnamefont {M.}~\bibnamefont {Schmidt}}, \bibinfo {author} {\bibfnamefont {C.}~\bibnamefont {Felser}}, \bibinfo {author} {\bibfnamefont {J.}~\bibnamefont {Wosnitza}},\ and\ \bibinfo {author} {\bibfnamefont {J.}~\bibnamefont {Gooth}},\ }\href {https://doi.org/10.1103/PhysRevB.102.035132} {\bibfield  {journal} {\bibinfo  {journal} {Physical Review B}\ }\textbf {\bibinfo {volume} {102}},\ \bibinfo {pages} {035132} (\bibinfo {year} {2020})}\BibitemShut {NoStop}%
\bibitem [{\citenamefont {Shekhar}\ \emph {et~al.}(2015)\citenamefont {Shekhar}, \citenamefont {Nayak}, \citenamefont {Sun}, \citenamefont {Schmidt}, \citenamefont {Nicklas}, \citenamefont {Leermakers}, \citenamefont {Zeitler}, \citenamefont {Skourski}, \citenamefont {Wosnitza}, \citenamefont {Liu}, \citenamefont {Chen}, \citenamefont {Schnelle}, \citenamefont {Borrmann}, \citenamefont {Grin}, \citenamefont {Felser},\ and\ \citenamefont {Yan}}]{shekhar_extremely_2015}%
  \BibitemOpen
  \bibfield  {author} {\bibinfo {author} {\bibfnamefont {C.}~\bibnamefont {Shekhar}}, \bibinfo {author} {\bibfnamefont {A.~K.}\ \bibnamefont {Nayak}}, \bibinfo {author} {\bibfnamefont {Y.}~\bibnamefont {Sun}}, \bibinfo {author} {\bibfnamefont {M.}~\bibnamefont {Schmidt}}, \bibinfo {author} {\bibfnamefont {M.}~\bibnamefont {Nicklas}}, \bibinfo {author} {\bibfnamefont {I.}~\bibnamefont {Leermakers}}, \bibinfo {author} {\bibfnamefont {U.}~\bibnamefont {Zeitler}}, \bibinfo {author} {\bibfnamefont {Y.}~\bibnamefont {Skourski}}, \bibinfo {author} {\bibfnamefont {J.}~\bibnamefont {Wosnitza}}, \bibinfo {author} {\bibfnamefont {Z.}~\bibnamefont {Liu}}, \bibinfo {author} {\bibfnamefont {Y.}~\bibnamefont {Chen}}, \bibinfo {author} {\bibfnamefont {W.}~\bibnamefont {Schnelle}}, \bibinfo {author} {\bibfnamefont {H.}~\bibnamefont {Borrmann}}, \bibinfo {author} {\bibfnamefont {Y.}~\bibnamefont {Grin}}, \bibinfo {author} {\bibfnamefont {C.}~\bibnamefont {Felser}},\ and\ \bibinfo {author} {\bibfnamefont {B.}~\bibnamefont
  {Yan}},\ }\href {https://doi.org/10.1038/nphys3372} {\bibfield  {journal} {\bibinfo  {journal} {Nature Physics}\ }\textbf {\bibinfo {volume} {11}},\ \bibinfo {pages} {645} (\bibinfo {year} {2015})},\ \bibinfo {note} {publisher: Nature Publishing Group}\BibitemShut {NoStop}%
\bibitem [{\citenamefont {Klotz}\ \emph {et~al.}(2016)\citenamefont {Klotz}, \citenamefont {Wu}, \citenamefont {Shekhar}, \citenamefont {Sun}, \citenamefont {Schmidt}, \citenamefont {Nicklas}, \citenamefont {Baenitz}, \citenamefont {Uhlarz}, \citenamefont {Wosnitza}, \citenamefont {Felser},\ and\ \citenamefont {Yan}}]{klotz_quantum_2016}%
  \BibitemOpen
  \bibfield  {author} {\bibinfo {author} {\bibfnamefont {J.}~\bibnamefont {Klotz}}, \bibinfo {author} {\bibfnamefont {S.-C.}\ \bibnamefont {Wu}}, \bibinfo {author} {\bibfnamefont {C.}~\bibnamefont {Shekhar}}, \bibinfo {author} {\bibfnamefont {Y.}~\bibnamefont {Sun}}, \bibinfo {author} {\bibfnamefont {M.}~\bibnamefont {Schmidt}}, \bibinfo {author} {\bibfnamefont {M.}~\bibnamefont {Nicklas}}, \bibinfo {author} {\bibfnamefont {M.}~\bibnamefont {Baenitz}}, \bibinfo {author} {\bibfnamefont {M.}~\bibnamefont {Uhlarz}}, \bibinfo {author} {\bibfnamefont {J.}~\bibnamefont {Wosnitza}}, \bibinfo {author} {\bibfnamefont {C.}~\bibnamefont {Felser}},\ and\ \bibinfo {author} {\bibfnamefont {B.}~\bibnamefont {Yan}},\ }\href {https://doi.org/10.1103/PhysRevB.93.121105} {\bibfield  {journal} {\bibinfo  {journal} {Physical Review B}\ }\textbf {\bibinfo {volume} {93}},\ \bibinfo {pages} {121105} (\bibinfo {year} {2016})}\BibitemShut {NoStop}%
\bibitem [{\citenamefont {Lee}\ \emph {et~al.}(2015)\citenamefont {Lee}, \citenamefont {Xu}, \citenamefont {Huang}, \citenamefont {Sanchez}, \citenamefont {Belopolski}, \citenamefont {Chang}, \citenamefont {Bian}, \citenamefont {Alidoust}, \citenamefont {Zheng}, \citenamefont {Neupane}, \citenamefont {Wang}, \citenamefont {Bansil}, \citenamefont {Hasan},\ and\ \citenamefont {Lin}}]{lee_fermi_2015}%
  \BibitemOpen
  \bibfield  {author} {\bibinfo {author} {\bibfnamefont {C.-C.}\ \bibnamefont {Lee}}, \bibinfo {author} {\bibfnamefont {S.-Y.}\ \bibnamefont {Xu}}, \bibinfo {author} {\bibfnamefont {S.-M.}\ \bibnamefont {Huang}}, \bibinfo {author} {\bibfnamefont {D.~S.}\ \bibnamefont {Sanchez}}, \bibinfo {author} {\bibfnamefont {I.}~\bibnamefont {Belopolski}}, \bibinfo {author} {\bibfnamefont {G.}~\bibnamefont {Chang}}, \bibinfo {author} {\bibfnamefont {G.}~\bibnamefont {Bian}}, \bibinfo {author} {\bibfnamefont {N.}~\bibnamefont {Alidoust}}, \bibinfo {author} {\bibfnamefont {H.}~\bibnamefont {Zheng}}, \bibinfo {author} {\bibfnamefont {M.}~\bibnamefont {Neupane}}, \bibinfo {author} {\bibfnamefont {B.}~\bibnamefont {Wang}}, \bibinfo {author} {\bibfnamefont {A.}~\bibnamefont {Bansil}}, \bibinfo {author} {\bibfnamefont {M.~Z.}\ \bibnamefont {Hasan}},\ and\ \bibinfo {author} {\bibfnamefont {H.}~\bibnamefont {Lin}},\ }\href {https://doi.org/10.1103/PhysRevB.92.235104} {\bibfield  {journal} {\bibinfo  {journal} {Physical Review B}\
  }\textbf {\bibinfo {volume} {92}},\ \bibinfo {pages} {235104} (\bibinfo {year} {2015})}\BibitemShut {NoStop}%
\bibitem [{\citenamefont {Balduini}\ \emph {et~al.}(2024{\natexlab{a}})\citenamefont {Balduini}, \citenamefont {Molinari}, \citenamefont {Rocchino}, \citenamefont {Hasse}, \citenamefont {Felser}, \citenamefont {Zota}, \citenamefont {Schmid},\ and\ \citenamefont {Gotsmann}}]{balduini_origin_2024}%
  \BibitemOpen
  \bibfield  {author} {\bibinfo {author} {\bibfnamefont {F.}~\bibnamefont {Balduini}}, \bibinfo {author} {\bibfnamefont {A.}~\bibnamefont {Molinari}}, \bibinfo {author} {\bibfnamefont {L.}~\bibnamefont {Rocchino}}, \bibinfo {author} {\bibfnamefont {V.}~\bibnamefont {Hasse}}, \bibinfo {author} {\bibfnamefont {C.}~\bibnamefont {Felser}}, \bibinfo {author} {\bibfnamefont {C.}~\bibnamefont {Zota}}, \bibinfo {author} {\bibfnamefont {H.}~\bibnamefont {Schmid}},\ and\ \bibinfo {author} {\bibfnamefont {B.}~\bibnamefont {Gotsmann}},\ }\href {https://doi.org/10.1103/PhysRevB.109.045148} {\bibfield  {journal} {\bibinfo  {journal} {Physical Review B}\ }\textbf {\bibinfo {volume} {109}},\ \bibinfo {pages} {045148} (\bibinfo {year} {2024}{\natexlab{a}})}\BibitemShut {NoStop}%
\bibitem [{\citenamefont {Balduini}\ \emph {et~al.}(2024{\natexlab{b}})\citenamefont {Balduini}, \citenamefont {Rocchino}, \citenamefont {Molinari}, \citenamefont {Paul}, \citenamefont {Mariani}, \citenamefont {Hasse}, \citenamefont {Felser}, \citenamefont {Zota}, \citenamefont {Schmid},\ and\ \citenamefont {Gotsmann}}]{balduini_probing_2024}%
  \BibitemOpen
  \bibfield  {author} {\bibinfo {author} {\bibfnamefont {F.}~\bibnamefont {Balduini}}, \bibinfo {author} {\bibfnamefont {L.}~\bibnamefont {Rocchino}}, \bibinfo {author} {\bibfnamefont {A.}~\bibnamefont {Molinari}}, \bibinfo {author} {\bibfnamefont {T.}~\bibnamefont {Paul}}, \bibinfo {author} {\bibfnamefont {G.}~\bibnamefont {Mariani}}, \bibinfo {author} {\bibfnamefont {V.}~\bibnamefont {Hasse}}, \bibinfo {author} {\bibfnamefont {C.}~\bibnamefont {Felser}}, \bibinfo {author} {\bibfnamefont {C.}~\bibnamefont {Zota}}, \bibinfo {author} {\bibfnamefont {H.}~\bibnamefont {Schmid}},\ and\ \bibinfo {author} {\bibfnamefont {B.}~\bibnamefont {Gotsmann}},\ }\href {https://doi.org/10.1103/PhysRevLett.133.096601} {\bibfield  {journal} {\bibinfo  {journal} {Physical Review Letters}\ }\textbf {\bibinfo {volume} {133}},\ \bibinfo {pages} {096601} (\bibinfo {year} {2024}{\natexlab{b}})}\BibitemShut {NoStop}%
\bibitem [{\citenamefont {Bachmann}\ \emph {et~al.}(2017)\citenamefont {Bachmann}, \citenamefont {Nair}, \citenamefont {Flicker}, \citenamefont {Ilan}, \citenamefont {Meng}, \citenamefont {Ghimire}, \citenamefont {Bauer}, \citenamefont {Ronning}, \citenamefont {Analytis},\ and\ \citenamefont {Moll}}]{bachmann_inducing_2017}%
  \BibitemOpen
  \bibfield  {author} {\bibinfo {author} {\bibfnamefont {M.~D.}\ \bibnamefont {Bachmann}}, \bibinfo {author} {\bibfnamefont {N.}~\bibnamefont {Nair}}, \bibinfo {author} {\bibfnamefont {F.}~\bibnamefont {Flicker}}, \bibinfo {author} {\bibfnamefont {R.}~\bibnamefont {Ilan}}, \bibinfo {author} {\bibfnamefont {T.}~\bibnamefont {Meng}}, \bibinfo {author} {\bibfnamefont {N.~J.}\ \bibnamefont {Ghimire}}, \bibinfo {author} {\bibfnamefont {E.~D.}\ \bibnamefont {Bauer}}, \bibinfo {author} {\bibfnamefont {F.}~\bibnamefont {Ronning}}, \bibinfo {author} {\bibfnamefont {J.~G.}\ \bibnamefont {Analytis}},\ and\ \bibinfo {author} {\bibfnamefont {P.~J.~W.}\ \bibnamefont {Moll}},\ }\href {https://doi.org/10.1126/sciadv.1602983} {\bibfield  {journal} {\bibinfo  {journal} {Science Advances}\ }\textbf {\bibinfo {volume} {3}},\ \bibinfo {pages} {e1602983} (\bibinfo {year} {2017})},\ \bibinfo {note} {number: 5 Publisher: American Association for the Advancement of Science}\BibitemShut {NoStop}%
\bibitem [{\citenamefont {Fuchser}(1971)}]{fuchser_anisotropic_1971}%
  \BibitemOpen
  \bibfield  {author} {\bibinfo {author} {\bibfnamefont {T.~D.}\ \bibnamefont {Fuchser}},\ }\href {https://digital.library.unt.edu/ark:/67531/metadc278322/} {{\selectlanguage {English}\bibinfo {title} {Anisotropic {Relaxation} {Time} for {Solids} with {Ellipsoidal} {Fermi} {Surfaces}}}} (\bibinfo {year} {1971}),\ \bibinfo {note} {publisher: North Texas State University}\BibitemShut {NoStop}%
\bibitem [{\citenamefont {Mackey}\ and\ \citenamefont {Sybert}(1968)}]{mackey_electron_1968}%
  \BibitemOpen
  \bibfield  {author} {\bibinfo {author} {\bibfnamefont {H.~J.}\ \bibnamefont {Mackey}}\ and\ \bibinfo {author} {\bibfnamefont {J.~R.}\ \bibnamefont {Sybert}},\ }\href {https://doi.org/10.1103/PhysRev.172.603} {\bibfield  {journal} {\bibinfo  {journal} {Physical Review}\ }\textbf {\bibinfo {volume} {172}},\ \bibinfo {pages} {603} (\bibinfo {year} {1968})},\ \bibinfo {note} {publisher: American Physical Society}\BibitemShut {NoStop}%
\bibitem [{\citenamefont {Price}(1960)}]{price_anisotropic_1960}%
  \BibitemOpen
  \bibfield  {author} {\bibinfo {author} {\bibfnamefont {P.~J.}\ \bibnamefont {Price}},\ }\href {https://doi.org/10.1147/rd.42.0152} {\bibfield  {journal} {\bibinfo  {journal} {IBM Journal of Research and Development}\ }\textbf {\bibinfo {volume} {4}},\ \bibinfo {pages} {152} (\bibinfo {year} {1960})},\ \bibinfo {note} {conference Name: IBM Journal of Research and Development}\BibitemShut {NoStop}%
\bibitem [{\citenamefont {Parrott}(1965)}]{parrott_new_1965}%
  \BibitemOpen
  \bibfield  {author} {\bibinfo {author} {\bibfnamefont {J.~E.}\ \bibnamefont {Parrott}},\ }\href {https://doi.org/10.1088/0370-1328/85/6/312} {\bibfield  {journal} {\bibinfo  {journal} {Proceedings of the Physical Society}\ }\textbf {\bibinfo {volume} {85}},\ \bibinfo {pages} {1143} (\bibinfo {year} {1965})}\BibitemShut {NoStop}%
\bibitem [{\citenamefont {Ham}\ and\ \citenamefont {Mattis}(1960)}]{ham_electrical_1960}%
  \BibitemOpen
  \bibfield  {author} {\bibinfo {author} {\bibfnamefont {F.~S.}\ \bibnamefont {Ham}}\ and\ \bibinfo {author} {\bibfnamefont {D.~C.}\ \bibnamefont {Mattis}},\ }\href {https://doi.org/10.1147/rd.42.0143} {\bibfield  {journal} {\bibinfo  {journal} {IBM Journal of Research and Development}\ }\textbf {\bibinfo {volume} {4}},\ \bibinfo {pages} {143} (\bibinfo {year} {1960})},\ \bibinfo {note} {conference Name: IBM Journal of Research and Development}\BibitemShut {NoStop}%
\bibitem [{\citenamefont {Englman}\ and\ \citenamefont {Sondheimer}(1956)}]{englman_electrical_1956}%
  \BibitemOpen
  \bibfield  {author} {\bibinfo {author} {\bibfnamefont {R.}~\bibnamefont {Englman}}\ and\ \bibinfo {author} {\bibfnamefont {E.~H.}\ \bibnamefont {Sondheimer}},\ }\href {https://doi.org/10.1088/0370-1301/69/4/304} {\bibfield  {journal} {\bibinfo  {journal} {Proceedings of the Physical Society. Section B}\ }\textbf {\bibinfo {volume} {69}},\ \bibinfo {pages} {449} (\bibinfo {year} {1956})}\BibitemShut {NoStop}%
\bibitem [{\citenamefont {Moll}(2018)}]{moll_focused_2018}%
  \BibitemOpen
  \bibfield  {author} {\bibinfo {author} {\bibfnamefont {P.~J.}\ \bibnamefont {Moll}},\ }\href {https://doi.org/10.1146/annurev-conmatphys-033117-054021} {\bibfield  {journal} {\bibinfo  {journal} {Annual Review of Condensed Matter Physics}\ }\textbf {\bibinfo {volume} {9}},\ \bibinfo {pages} {147} (\bibinfo {year} {2018})},\ \bibinfo {note} {number: 1 \_eprint: https://doi.org/10.1146/annurev-conmatphys-033117-054021}\BibitemShut {NoStop}%
\end{thebibliography}


\end{document}